%% file: HighEffMuonRegSys.tex
\documentclass[review]{elsarticle}
\usepackage[utf8]{inputenc}  
\usepackage{lineno}
\usepackage{hyperref}
\hypersetup{
	colorlinks,
	citecolor=blue,
	filecolor=blue,
	linkcolor=blue,
	urlcolor=blue,
}
\usepackage{subfig}
\usepackage{textcomp} 
\usepackage{xcolor}
\usepackage{ulem}
\usepackage{amsmath}
\usepackage{amsfonts}
\usepackage{amssymb}
\usepackage{epsfig}

\usepackage{graphicx}

\biboptions{sort&compress}

\journal{Journal of Nucl. Instrum. Methods Phys. Res. A}

\begin{document}
	
	\begin{frontmatter}
		
		\title{High efficiency muon registration system based on scintillator strips}
		\author[JINR,STU]{A.~Artikov}
		\author[JINR]{V.~Baranov}
		\author[JINR]{A.~Boikov}
		\author[JINR,GTU]{D.~Chokheli} \corref{mycorrespondingauthor} 
		\cortext[mycorrespondingauthor]{Corresponding author}
		\ead{Davit.Chokheli@cern.ch}
		\author[JINR]{Yu.I.~Davydov}
		\author[JINR]{V.~Glagolev}
		\author[JINR]{A.~Simonenko}
		\author[JINR,GTU]{Z.~Tsamalaidze}
		\author[JINR]{I.~Vasilyev}
		\author[JINR]{I.~Zimin}
		
		\address[JINR]{Joint Institute for Nuclear Research, Dubna, Russia}
		\address[STU]{Nuclear Physics Laboratory, Samarkand State University, Samarkand, Uzbekistan}
		\address[GTU]{Georgian Technical University, Tbilisi, Georgia}

		\begin{abstract}
			\input{parts/Abstract}

		\end{abstract}


	\end{frontmatter}
	\pagebreak
	\tableofcontents
	\pagebreak
	
	\section{Introduction}
	\input{parts/Introduction.tex}\label{introduction}
	
	\section{Theoretical basement for calculation of the efficiency of Cosmic Ray Veto (CRV) system}\label{theory}
	\input{parts/Theory.tex}
	
	\section{The simplified method for light yield simulation}\label{transscan}
	\input{parts/SimpMeth.tex}

	\section{Simulation of the CRV module charged particle registration probability by GEANT-4}\label{regprob}
	\input{parts/SimulationModule.tex}
	
	\section{Experimental study of the 4x4 CRV module prototype with cosmic muons}\label{expeffmodule}
	\input{parts/ExpStudy4x4.tex}

	\section{Conclusions}\label{conclusions}
	\input{parts/Conclusions.tex}

	\section{Acknowledgments}\label{thanks}
	\input{parts/Acknowledgments.tex}

	
    \bibliography{
    	mybib.bib
  }

\end{document}

%% file: parts/Abstract.tex
Experiments such as mu2e (FNAL, USA) and COMET (KEK, Japan), seeking the direct muon-to-electron conversion as part of the study of Charged Leptons Flavor Violation processes, should have a extremely high, up-to 99.99\%, efficiency muon detection system with a view to their subsequent suppression as background. In this article, the possibility to achieve such efficiency for a short and long term is discussed for modules based on 7- or 10-mm-thick but same 40-mm-wide plastic scintillation strips with single 1.2 mm WLS fiber glued into the groove along the strip and using MPPC/SiPM for light detection.

A Simplified Light Yield Distribution method to estimate the efficiency of the module was proposed and the simulation results obtained with GEANT 4 for a system based on a 4-by-4 array of 7x40x3000 mm strips compared with the experimental data. Found that for the systems required the high level registration efficiency at the 99.99\% and more, it is important to improve the light yield as much as possible and achieve the gap between neighbor scintillation volumes as small as possible.

%% file: parts/Introduction.tex
Detectors based on a plastic scintillator were widely used in most modern experiments in High Energy Physics. Typically, the detection efficiency of minimum ionizing particles for such systems is sufficient to be more than 90\%. 

However, some experiments need to have so-called active shield system from background muons, particularly for cosmic muons. This means that the cosmic muon should first be detected and then rejected through online/offline data processing. We call such a system as a Cosmic Ray Veto (CRV) system. 

CRV system always requires a higher efficiency for muon registration compared to regular muon systems. For instance, both the Mu2e experiment \cite{mu2e_tdr2014} (FNAL, USA) and the COMET experiment \cite{comet_tdr2018} (KEK, Japan) require that the CRV system registration efficiency be on the 99.99\% level to establish the sufficient suppression of the cosmic background and thus achieve the required sensitivity of these experiments in order of on $10^{-17}$ for a so called single-event – for a direct conversion of the muon into the electron. 

%% file: parts/Theory.tex
Usually, the muon registration system is the hodoscope/array of the several layers of plastic scintillation bars and/or drift/proportional/resistive plate chambers and could be split into the modules. In this article, we are considering a 4-layer CRV module based on 40-mm-wide plastic scintillation strips with the same layout but with different strips thickness: of 7-mm-thick the first one and 10-mm-thick the second.

\subsection{Charged particle registration probability for 4-layer CRV module}\label{secprobCRV}
Considering CRV module is an array of the four layers of the strips with 16 strips on each layer, and each layer is identical to other and requesting the coincidence of any 3 layers of 4, - with such conditions we can estimate the registration efficiency for charged particle of this system using combinatorial mathematics as follow:

\begin{equation} \label{eq:prop3of4_ideal}
P_{CRV} = \textrm{C}_{4}^{3}\times (P_L)^{3}\times(1-P_L)+(P_L)^{4}
\end{equation}
here: $\textrm{C}_{4}^{3}$ could be calculated as: $\textrm{C}_{4}^{3} = \frac{4!}{(4-3)!3!} = 4 $

Using this equation (1) and in case of efficiency for CRV module on charged particles registration is required to be on a level of $P_{CRV} =99.99\%$, the efficiency of each layer should be better than  $P_L = 99.65\% $.

Unfortunately, the efficiency of each layer is not a constant one and varies from case to case due to different factors (production quality, components properties variation, etc.). So, to calculate the registration probability for the real module, we should use the equation, which includes variation of registration probability of each strip fired:

\begin{equation} \label{eq:prop3of4_real}
\begin{split}
P_{m} = \sum \limits_{n=0}^{3}P_{L(i\%4)}\times P_{L((i+1)\%4)}\times P_{L((i+2)\%4)}\times (1-P_{L((i+3)\%4)}) + \\
+ P_{L0}\times P_{L1}\times P_{L2}\times P_{L3}
\end{split}
\end{equation}

In this equation, $P_{L0}; P_{L1}; P_{L2}; P_{L3}$ are the registration probability for layer 1, 2, 3 and 4 respectively\footnote{here: $
\sum \limits_{n=0}^{3}P_{L(i\%4)}\times P_{L((i+1)\%4)}\times P_{L((i+2)\%4)}\times (1-P_{L((i+3)\%4)}) = \\
= P_{L0}\times P_{L1}\times P_{L2}\times (1-P_{L3}) + P_{L0}\times P_{L1}\times (1-P_{L2})\times P_{L3} + \\
+ P_{L0}\times (1-P_{L1})\times P_{L2}\times P_{L3} + (1-P_{L0})\times P_{L1}\times P_{L2}\times Prob_{L3}$}.

\subsection{Combined efficiency for one layer consisted with N-strips}
Each layer of CRV module represents a 1D-array of plastic scintillation strips. Since cosmic muon could pass multiple stripes depending on its incidence angle to the module surface, we need to combine the neighbor strips registration probabilities to each other in aim to calculate in proper way each layer registration probability. To facilitate the calculation and make it simple, we can calculate inefficiency $\overline{P_{L}}$ (the probability to miss detection of the particles passage) of each strip first and then we can find the inefficiency of layer as a multiplication of registration inefficiency of the strips with follow equation: $
\overline{P_{L}} = \overline{P_{S_1}}\times\overline{P_{S_2}}\times\ldots\times\overline{P_{S_N}} = \prod\limits_{i=1}^{N}\overline{P_{S_i}}$ (here: $\overline{P_{S_1}}$, $\overline{P_{S_2}}$, $\overline{P_{S_N}}$ - inefficiency for strip 1, strip 2, strip N respectively). And now we can easily calculate the registration probability for the layer as:

\begin{equation} \label{eq:LayerEff}
\begin{split}
P_{L} = 1 - \overline{P_{L}} = 1 - \prod \limits_{i=1}^{N}\overline{P_{S_i}} = 1 -  \prod \limits_{i=1}^{N}(1-P_{S_i})
\end{split}
\end{equation}

For instance, if a layer consists of 16 strips and a charged particle passes through $5^{th}$ and $6^{th}$ neighbor strips only, so inefficiency for those strips could be calculated as $\overline{P_{5}} = (1-P_{(S_5)})$ and $\overline{P_{6}} = (1-P_{(S_6)})$ but the inefficiency for other strips should be equal to 1 since they never fired. And overall efficiency of this layer according to the equation \eqref{eq:LayerEff} could be expressed as 
$P_{L} = 1 - \overline{P_{L}} = 1 - (1-P_{(S_5)})(1-P_{(S_6)})$. Next step is to calculate the registration probability for each strip.

\subsection{Charged particle registration probability for the particular strip}
The values of the light yield collected on a photosensitive detector (PMT, SiPM/MPPC, etc.) randomizes around of some mean value and its probabilities can be calculated using Poisson distribution \cite{poissdiss}

\begin{equation} \label{eq:Poisson}
P(x) = \frac{\mu^{x}e^{-\mu}}{x!}
\end{equation}

here $x$ is amount of light yield collected on PMT or SiPM in photoelectrons and $\mu$ is expected/mean value of the collected on a photosensor light in photoelectrons.

The Poisson distribution for a high number of light yield value tends to a Gaussian distribution with $\sigma = \sqrt{\mu}$ and the distribution of probability for light yield values could be calculated as:

\begin{equation} \label{eq:Gauss}
G(x) = \frac{1}{\sqrt{2\pi\mu}}e^{-\frac{(x-\mu)^2}{2\mu}}
\end{equation}

Usually, to distinguish the real signal from the noise, the incoming signal should be processed by a discriminator with some level of threshold (by charge, by voltage amplitude of pulse, by current, etc.). But the pulse discrimination will suppress not only noise but some necessary events as well thus decreasing particle registration efficiency for the whole system. In this case, the probability of the registration for the charged particle passed the detector could be calculated as an integral from the discrimination threshold value to the infinitive [7]:

\begin{equation} \label{eq:GaussInt}
P(\mu) = \int_{T_{ph.e.}}^{+\infty}f_{eff}(x)dx = \frac{1}{\sqrt{2\pi\mu}}\int_{T_{ph.e.}}^{+\infty}e^{-\frac{(x-\mu)^2}{2\mu}}dx
\end{equation}

Of course, for minimum ionizing particles the more appropriate formalization is the Landay distribution. But, in some cases, for instance when just the number of events is important not the shape and the threshold value is less than a mean value (or, in other words the threshold cuts part of the left wing of the distribution), we can use Gaussian formalization with some approximation as a mimic of left wing of Landau distribution.

Therefore, to estimate registration probability, it is possible to use an error function ($erf(y) = \frac{2}{\sqrt{\pi}}\int_{0}^{y}e^{-x^2}dx)$) which is an anti-derivative of \eqref{eq:GaussInt} and the final equation to calculate registration probability of the charged particle passage through the strip should look like:

\begin{equation} \label{eq:ErfInt}
P(\mu) = \frac{erf\left( \frac{+\infty-\mu}{\sqrt{2\mu}}\right)}{2} - \frac{erf\left( \frac{T_{ph.e.}-\mu}{\sqrt{2\mu}}\right)}{2} = \frac{1}{2} + \frac{erf\left( \frac{\mu-T_{ph.e.}}{\sqrt{2\mu}}\right)}{2}
\end{equation}

For instance, if the mean value of the light yield collected on photodetector is 15.7 ph.e. and the threshold is set on 5 ph.e. level then the registration probability for the strip should be $P(\mu) = 99.65\% $. And it is the theoretical minimum amount of light in ph.e. from strip collected with selected threshold to satisfy required level of 99.99\% registration efficiency for 4-layer CRV system in condition of coincidence for any 3 layers of 4 (see \ref{secprobCRV}).

%% file: parts/SimpMeth.tex
In aim to predict the light yield value collected on photodetector, one can use a direct simulation of the propagation for a charged particle through scintillator body by counting each processes emitting the light. But this method consumes too much processor time for the calculation since of a huge secondary particle number and requires a very precise simulation model with correct coefficients for a particular scintillator type and requires very careful feedback checking. Therefore, we need to try to find simplified models with less consumption of the calculation time and less uncertainty with light yield calculation.

In this chapter we will discuss how to simplify the light yield estimation and thus drastically improve the time consumption needed for calculations. One of such ways is based on a study to find a direct dependence of the light yield on charged particle path which should be done for the scintillator strips we will use in a future. We will call it as “Simplified Light Yield Distribution” method (SLYD method).

\subsection{Strip transverse scan as an essential part of SLYD method}

This study is based on finding the dependence for the light yield per path unit (for instance, in photoelectron per mm) for real detector and then to use obtained results to predict the CRV overall efficiency with modeling in Geant-4 \cite{Giant4}. The main idea of this method is as follows.

\begin{figure}[!htbp]
    \centering
    \includegraphics[width=0.75\linewidth, keepaspectratio]{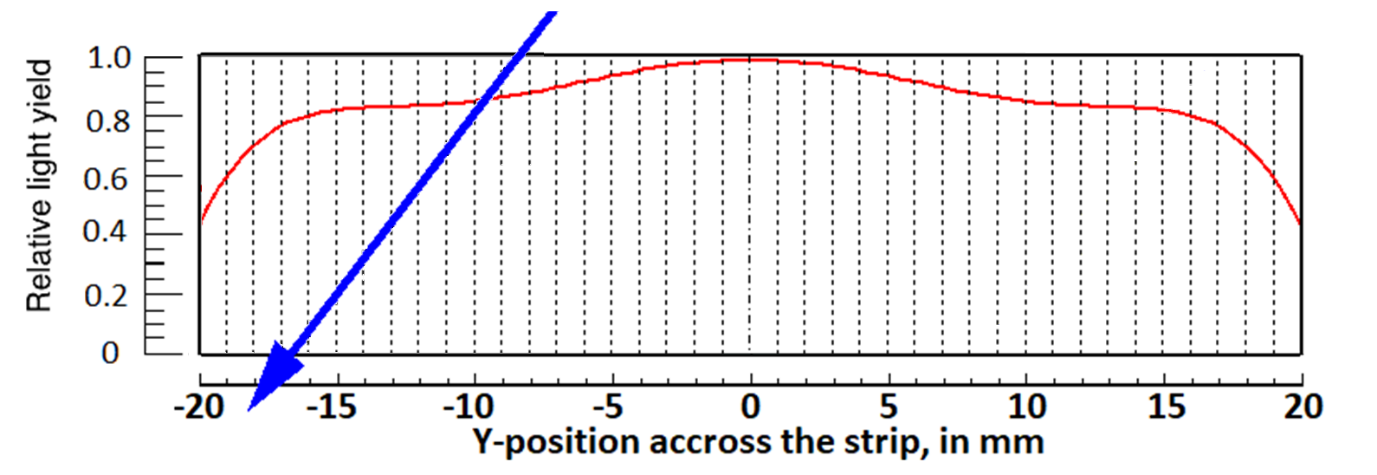}
    \caption{\small The cross-section of the strip sliced into areas split by dashed lines; a muon path (blue arrow) and a relative light yield distribution (red curve utilizing 6th degree polynomial function) found by transverse scan of the strips.}
    \label{fig: muonpath}
\end{figure}

As we know, charged particle creates some light while a passage through the scintillation strip, or, in other words, the light yield from scintillation strip depends on a particle’s passage path and where this path lies, for instance, the light collected on edges is less than on central area across of strip, also the amount of light collected increases with the length of the particle path. In first approximation, we can slide the propagation path into the separate areas and, thus, an overall light yield could be presented as a sum of light yields obtained in each passage area (Fig. \ref{fig: muonpath}). The blue arrow shows the muon path direction, and the red curve introduces a light yield distribution $F_{\mu}(y)$ across a real scintillation strip with WLS fiber in and obtained by fitting the experimental data. 

Another, but important assumption is the homogeneity of the light collection inside of the selected area and the precision of this model should largely depend on quality of the strip production. 

In general, the light yield varies in different areas. To find the light yield in each area (Fig. \ref{fig: muonpath}), the data were collected and analyzed for the particles entering the strip in selected area in perpendicular to the strip surface. In this case the particle will pass inside of one area and thus it could be possible to find the mean value of light yield for this area. Since passage path length in this case is roughly equal to the strips thickness – the light yield per mm should be easily calculated. Once the light yield for each area obtained, the total light yield could be calculated as follow:

\begin{equation} \label{eq:simlight}
	\begin{split}
		\mu = \int_{-w/2}^{+w/2}{F_{\mu}(y)} \approx \sum_{i=0}^{N}{\mu_i L_i}
	\end{split}
\end{equation}

Here, $w$ is a strips width, $F_{\mu}(y)$ is a light yield distribution by the path, N is a total number of the areas, $\mu_{i}$ is a light yield per mm inside of a particular area between two neighbor dashed lines of the strip and $L_{i}$ is the muon path length in mm inside of this area.

\subsection{Transverse scan with cosmic rays for 7 mm and 10 mm thick and 40 mm wide strips}

\begin{figure}[!htbp]
	\centering
	\subfloat[]{\includegraphics[height=0.15\textheight, keepaspectratio]
		{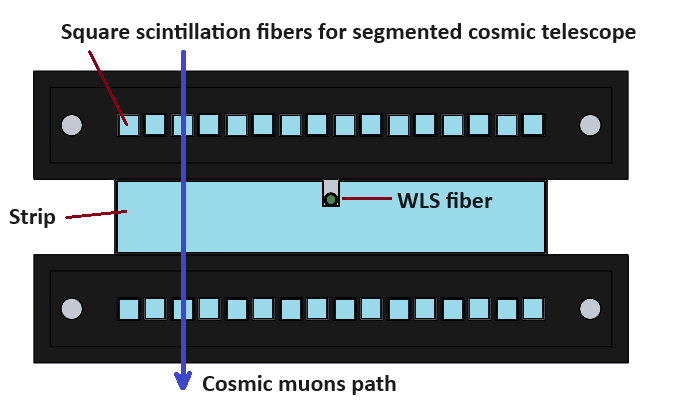} \label{fig: cosmicscansetup1}}
	\subfloat[]{\includegraphics[height=0.15\textheight, keepaspectratio]
		{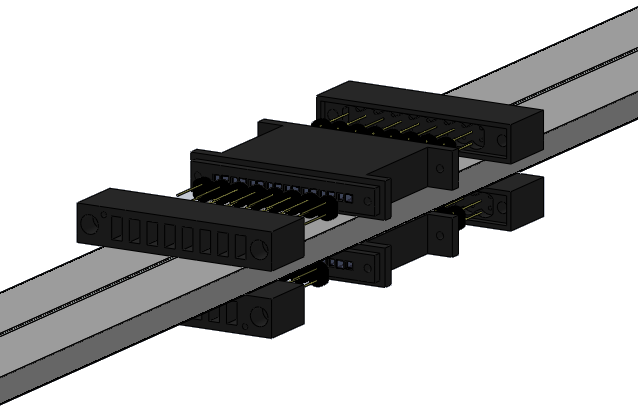} \label{fig: comsicscansetup2}}
	\caption{\small Scintillation strip with segmented cosmic muons telescope for the traverse scan(a) and layout of the experimental setup to provide the cosmic run with segmented telescope}
	\label{fig: cosmicscansetup}
\end{figure}

Long-term run with cosmic muons using segmented cosmic ray telescope could allow us to mimic the transverse scan of the strips (Fig. \ref{fig: cosmicscansetup1}). To provide it, the cosmic telescope with an active area of 40x40 mm2 was created. This telescope consisted of two 16-channel cosmic muon hodoscopes lying above and below testing strips (Fig. \ref{fig: comsicscansetup2}). Each channel is created by 2x2 $mm^2$ Kuraray SCSF-81J scintillating fiber \cite{kuraraySCSF81} placed with 2.5 mm pitch and directed along the strips. This geometry of hodoscope ensured to separate the telescope into 16 areas with step of 2.5 mm. The muon hodoscope was installed at 250 cm distance from strip readout by SiPMs. 

Four 3-m-long  scintillator strips for this study were produced by “Uniplast” (Vladimir, Russia \cite{uniplast}). Two of them had a cross-section of 7x40 $mm^2$ and other two had a cross-section of 10x40 $mm^2$. Each strip were equipped with a single WLS 1.2 mm WLS Kuraray Y11(200) fiber \cite{kurarayY11} glued to the groove (Fig. \ref{fig: cosmicscansetup1}). All strips were stacked vertically, and the telescope was positioned above and below the stack to ensure simultaneous cosmic run of the strip under study.

DAQ was based on CAEN DT5702 32-ch MPPC/SiPM readout Front-End \cite{caenDT5702}. Hamamatsu S13360-1350CS SiPMs \cite{hamamatsuS13360} with 1.3x1.3 $mm^2$ active surface were used for light collection from the strips and with the hodoscopes.

\begin{figure}[!htbp]
    \centering
    \includegraphics[width=0.75\linewidth, keepaspectratio]{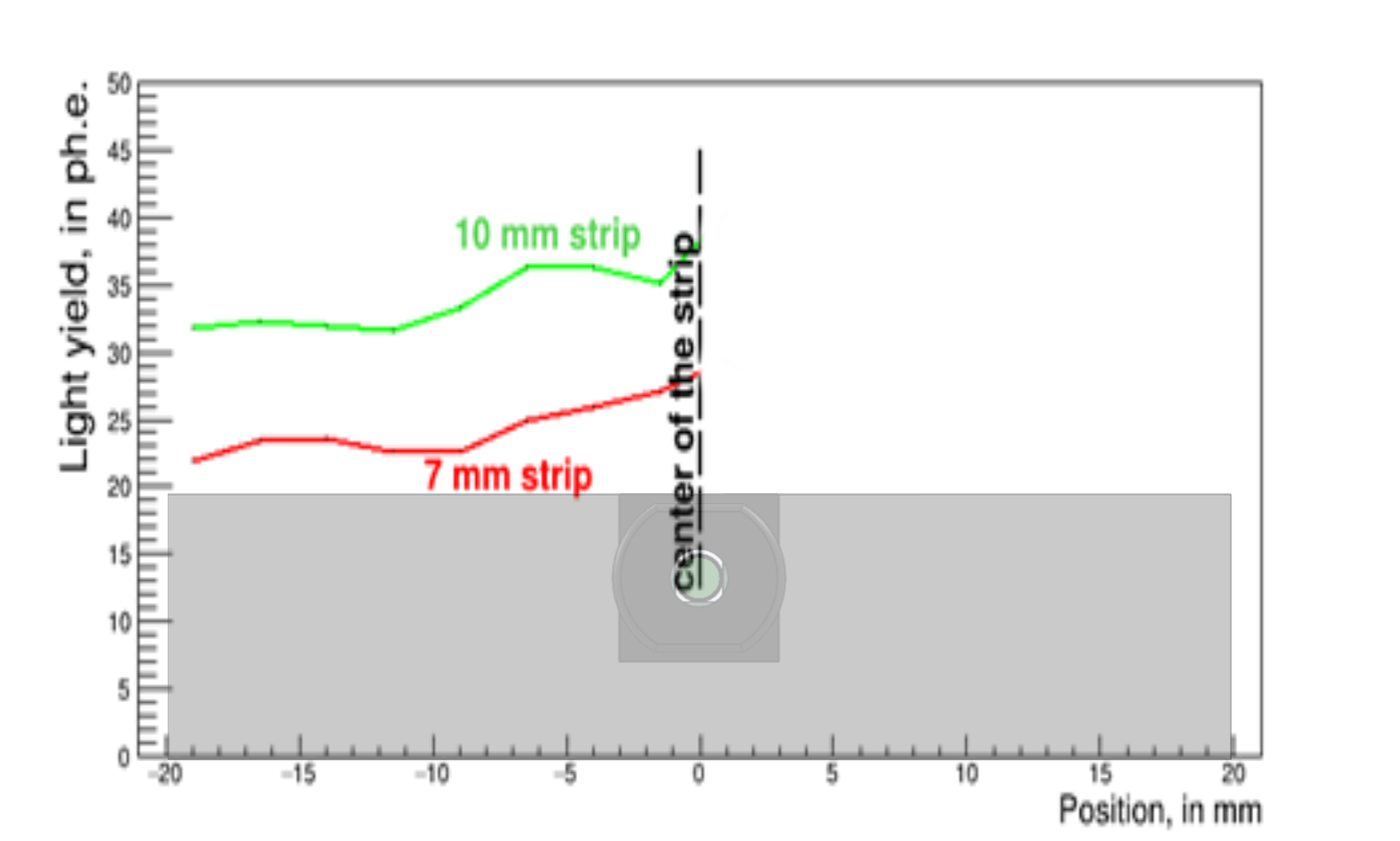}
    \caption{\small Cosmic run: average light yield (in photoelectrons) transverse distribution for 7-mm-thick and for 10-mm-thick strips}
    \label{fig: cosmicscanave}
\end{figure}

Data with cosmic muons was collected during of 2 weeks which allowed us to achieve about 150 “vertical” muons per position (corresponding hodoscope channels in coincidence, see Figure \ref{fig: cosmicscansetup1}). Light yield (in photoelectrons) transverse distribution for each strips is presented in Table \ref{table: cosresult} and averaged light yield distribution for each type of strips are presented in Figure \ref{fig: cosmicscanave}). Since we had limited number of SiPM on time for this test, only for half of strip was collected the cosmic data, having in the mind that other wing of strip should be symmetric to examined one. 

\begin{table}[!htb]
\fontsize{7pt}{7pt}\selectfont
    \centering
    \begin{tabular}{|l|l|l|l|l|l|l|l|l|}
    \hline
        Position, in mm & -19 & -16.5 & -14 & -11.5 & -9 & -6.5 & -4 & -1.5 \\ \hline
        7-mm-thick strip, \#1& 24.2 & 25.8 & 26.8 & 24.2 & 25.9 & 27.8 & 27.7 & 28.8 \\ \hline
        7-mm-thick strip, \#2& 19.5 & 20.9 & 20.3 & 20.9 & 19.2 & 21.9 & 24.3 & 25.4 \\ \hline
        10-mm-thick strip, \#3& 27.4 & 34.1 & 32 & 30 & 34.9 & 34.6 & 37.3 & 36.4 \\ \hline
        10-mm-thick strip, \#4& 36.3 & 30.5 & 32 & 33.1 & 31.8 & 38 & 35.3 & 33.9 \\ \hline
        “7 mm” average & 21.8 & 23.3 & 23.5 & 22.6 & 22.5 & 24.9 & 26 & 27.1 \\ \hline
        “10 mm” average & 31.8 & 32.3 & 32 & 31.6 & 33.3 & 36.3 & 36.3 & 35.2 \\ \hline
        ~ & edge of strip & ~ & ~ & ~ & ~ & ~ & ~ & Center \\ \hline
    \end{tabular}
    \caption{\small Cosmic run: light yield (in photoelectrons) transverse distribution for each strips}
    \label{table: cosresult}
\end{table}

\subsection{Strip transversal scan using collimated \texorpdfstring{$^{90}Sr/^{90}Y$}{Lg} \texorpdfstring{$\beta-$}{Lg}source}

One can see that precise of study on cosmic muons described in previous chapter is not enough to use obtained data in a future since low statistic and big pitch value. It is necessary to increase the statistic by order and to decrease pitch value from 2.5 mm to 0.5 mm at least to get necessary precision. The required improvement will drastically increase the difficulty of cosmic muon telescope structure and, more important, the time for data collection up to several months. 

\begin{figure}[!htbp]
	\centering
	\subfloat[]{\includegraphics[height=0.20\textheight, keepaspectratio]
		{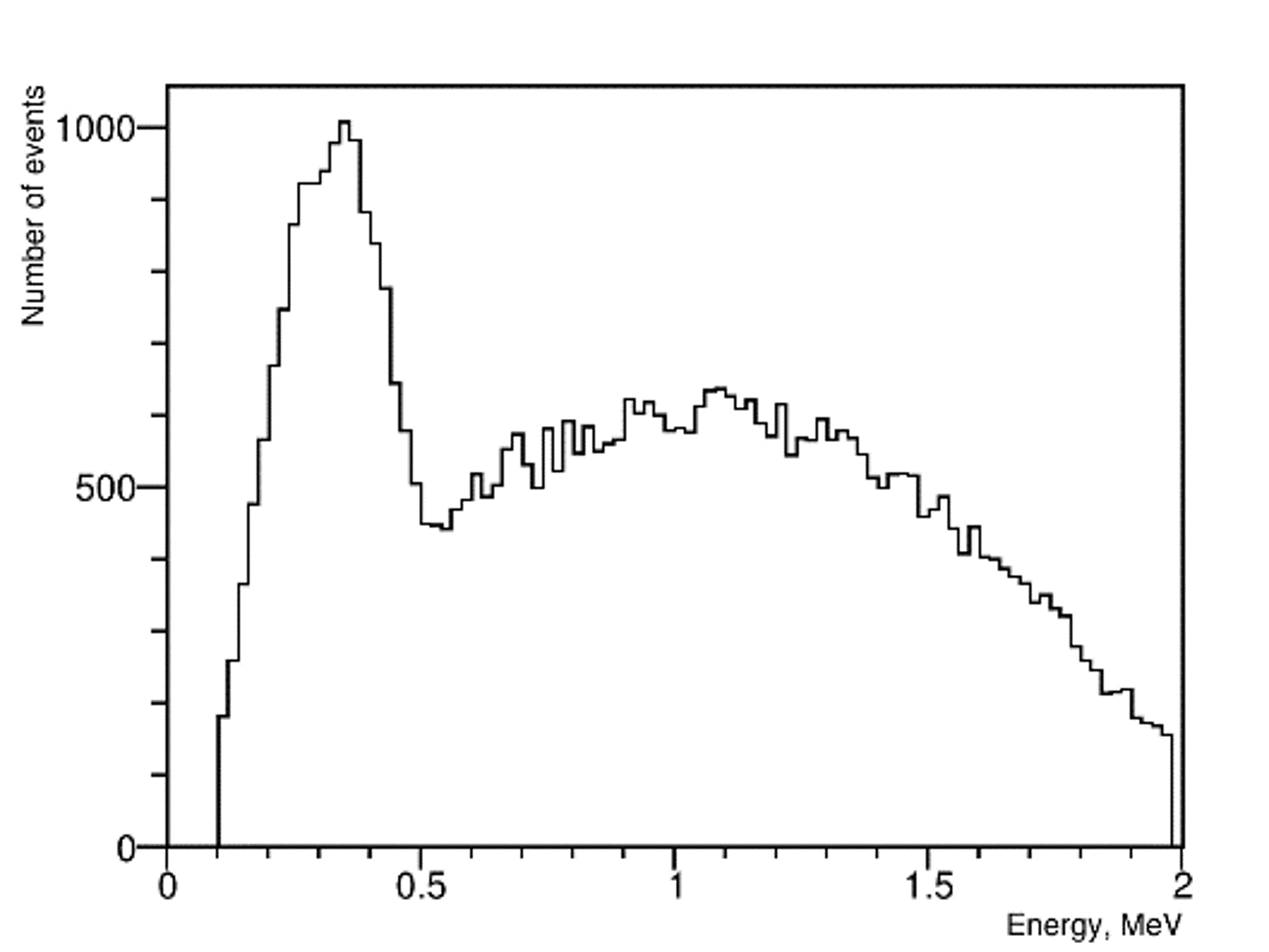} \label{fig: sourcenenergy}}
	\subfloat[]{\includegraphics[height=0.20\textheight, keepaspectratio]
		{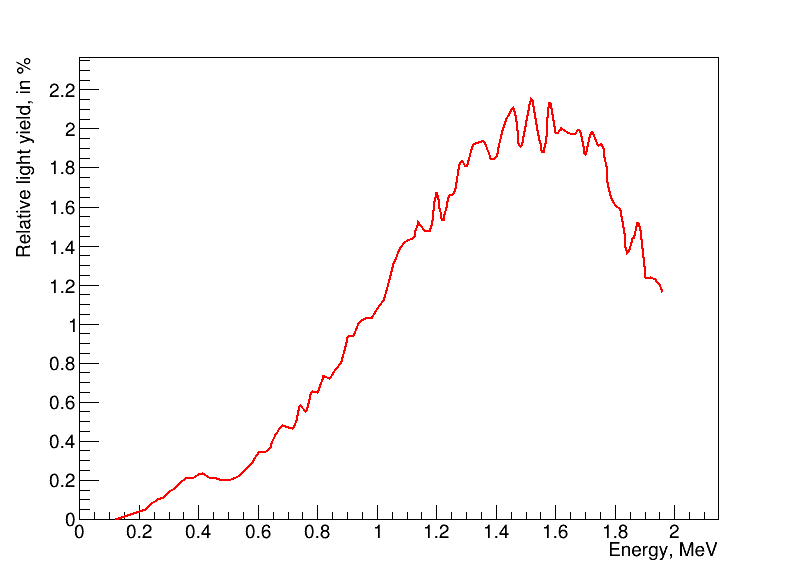} \label{fig: sourcerellight}}
	\caption{\small Simulation (done in Geant-4) of the energy distribution for $\beta-$particles beam issued by ${}^{90}Sr/{}^{90}Y$ source and collimated to 1 mm on output (a) and the distribution of relative light yield (in percent) as a function of the energy deposited to the strip (b)}
	\label{fig: sourcesim}
\end{figure}

The transverse scan of the strip with $\beta-$source mimicking the cosmic muon is another way to obtain necessary data in a reasonable time.
The $\beta-$particle beam issued by $^{90}Sr/^{90}Y$ source and collimated to 1 mm in diameter could be suitable for this purpose. To exam this proposal, we simulated a rate distribution by an energy for such beam (Fig. \ref{fig: sourcenenergy}). And then obtained distribution was transformed into to the distribution of relative energy deposition to the strip (which corresponds to the light yield collected on this strip) by $\beta-$particle energy (Fig. \ref{fig: sourcerellight}). Please note, that average $\beta-$particles energy for $^{90}Sr \rightarrow ^{90}Y$ decay is about 0.205 MeV and for $^{90}Y \rightarrow ^{90}Zr$ decay is about 0.93 MeV.

According to distribution of the energy deposition into a strip (Fig. \ref{fig: sourcerellight}) the most contribution to light yield is done by beta particles bigger than 1 MeV with maximum for 1.5 MeV particles. The radiation length for 1 MeV $\beta-$particles is about 3.6 mm and for 1.5 MeV – about 5.4 mm. So, for scintillators with less than 10 mm in thickness such source will somehow mimic the cosmic muons thus drastically speeding up the transverse scan relative to such test using cosmic rays. 

\begin{figure}[!htbp]
	\centering
	\subfloat[]{\includegraphics[height=0.20\textheight, keepaspectratio]
		{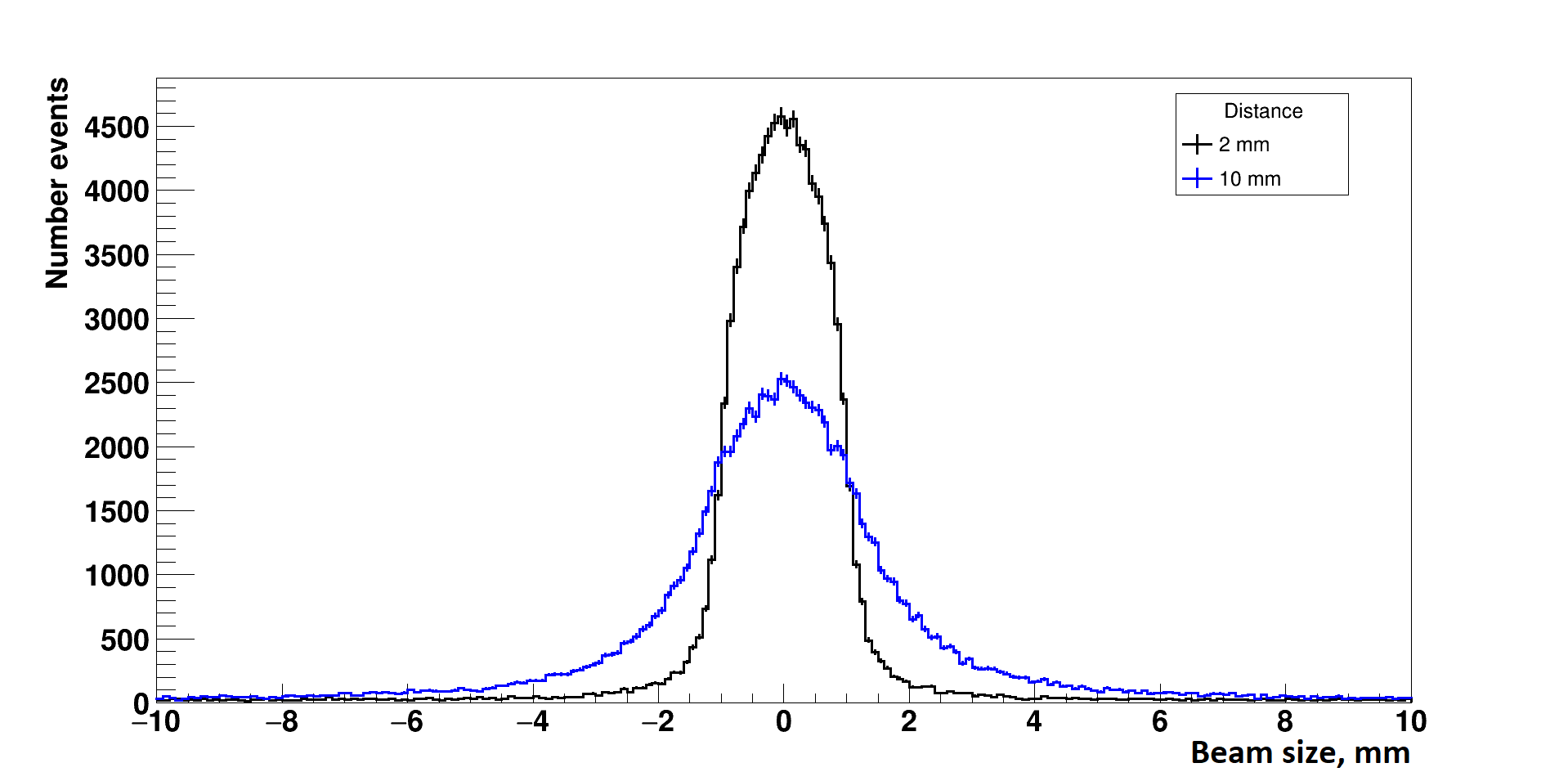}\label{fig: colbeamsim}}
	\subfloat[]{\includegraphics[height=0.20\textheight, keepaspectratio]
		{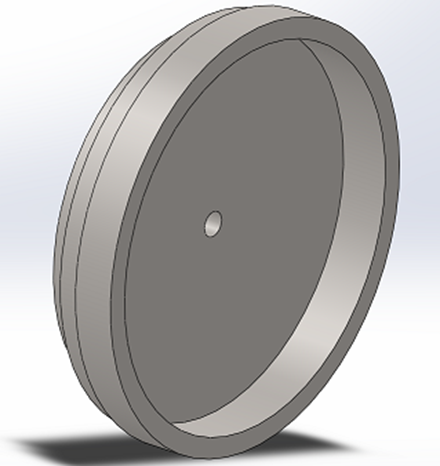} \label{fig: colbeamform}}
	\caption{\small Simulation of the beam size (by $^{90}Sr/^{90}Y$ $\beta-$source) with different distances after the 1 mm diameter collimator (a) and Al-collimator design (b)}
	\label{fig: colbeam}
\end{figure}

At first, we performed the simulation using GEANT-4 \cite{Giant4} environment for the $\beta-$particles beam divergence by distance between a collimator and a strip entrance surface (Fig. \ref{fig: colbeamsim}): for 2 mm distance (black curve) and for 10 mm (blue curve). The material for the collimator was an aluminum disk with a diameter of 10 cm and a thickness of 3 mm with a hole with a diameter of 1 mm in the center, aligned with the source.

\begin{figure}[!htbp]
	\centering
	\subfloat[]{\includegraphics[height=0.20\textheight, keepaspectratio]
		{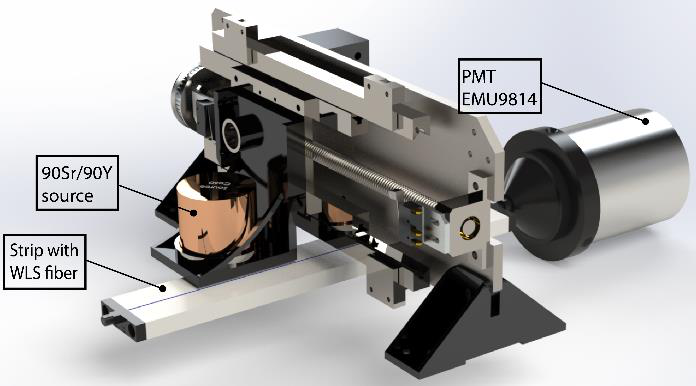}}
	\subfloat[]{\includegraphics[height=0.20\textheight, keepaspectratio]
		{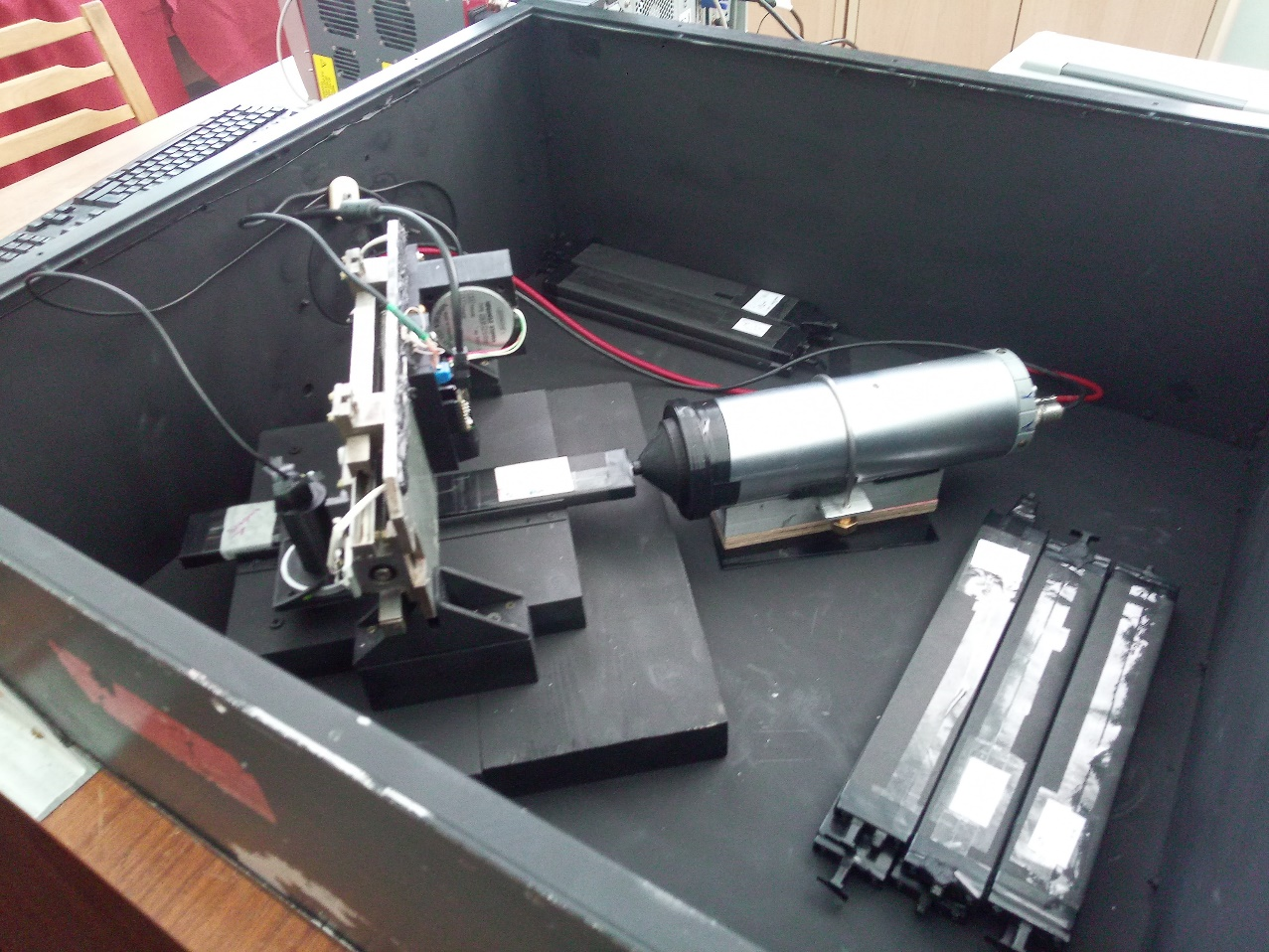}}
	\caption{\small Layout of the experimental setup to provide the transverse scan of the strip (a) and the real look of this setup (b)}
	\label{fig: scansetup}
\end{figure}

The special setup was built to provide such a transverse scan using $^{90}Sr/^{90}Y$ $\beta-$source with about 0.03 mCi activity (Fig. \ref{fig: scansetup}). The beam formed using collimator made from a 3-mm-thick aluminum disk with 1 mm-diameter in the center (Fig. \ref{fig: colbeamform}). The step of scanning was set to 0.5 mm. The light from strip was collected using EMI9814 PMT \cite{emi9814} and PMT’s anode current was measured with Keithley 6847 picoammeter \cite{picoamK6487}.

\subsection{Traverse scan results for 7 mm and 10 mm thick and 40 mm wide strips}

To obtain a distribution of light collection over the strip width, we scanned the strip across its width by collimated $\beta-$source as described in previous chapter. Two types of strips were scanned at this stage: 7x40 mm and 10x40 $mm^2$ strips in cross-section. These strips were also produced by Uniplast (Vladimir, Russia). Both had only one fiber 1.2mm WLS Kuraray Y11(200), double cladding, C-type \cite{kurarayY11} glued into the 2-mm-deep groove along the strip. 

We provided the scans for strips in two positions: when WLS-fiber was close to beam (on “top”) and then the strip was turned over (WLS-fiber is on “bottom”). These distributions for each strip (for “top” and “bottom”) then summarized into one plot and normalized to 1 and then fitted with 6$^{th}$ degree polynomial function to achieve the best approximation.

\begin{figure}[!htbp]
	\centering
	\subfloat[]{\includegraphics[height=0.1\textheight, keepaspectratio]
		{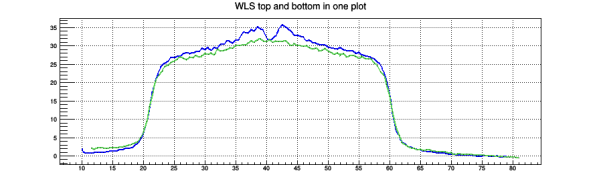}}
	\subfloat[]{\includegraphics[height=0.1\textheight, keepaspectratio]
		{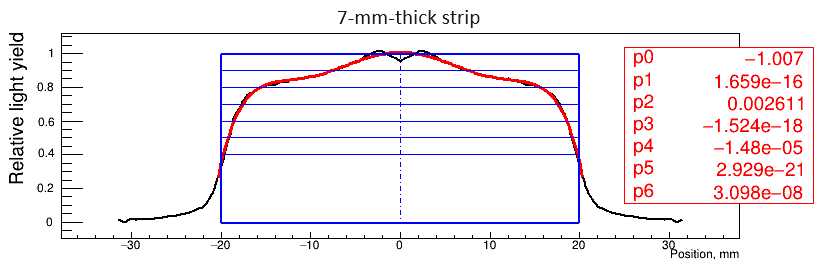}}
	\caption{\small The distributions of light yield across a for 7-mm-thick strip for “top” (blue) and “bottom” (green) directions (a). Normalized distribution (b) for summary of “top” and “bottom” distributions, a blue rectangular illustrates the strips geometry.}
	\label{fig: scanres07mm}
\end{figure}

\begin{figure}[!htbp]
	\centering
	\subfloat[]{\includegraphics[height=0.095\textheight, keepaspectratio]
		{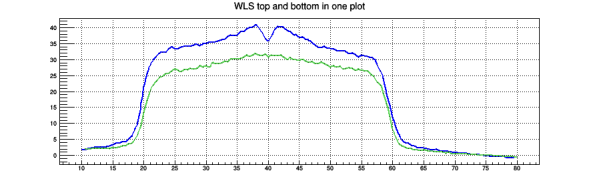}}
	\subfloat[]{\includegraphics[height=0.095\textheight, keepaspectratio]
		{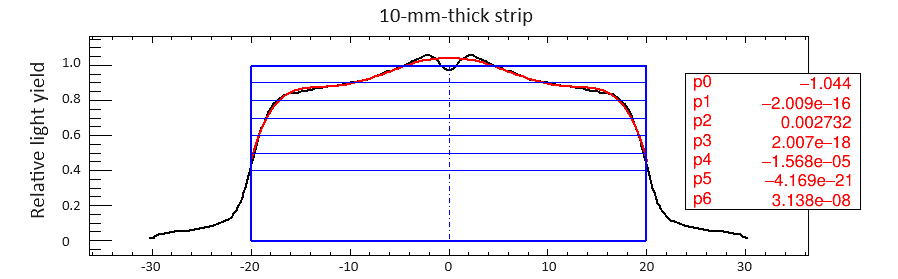}}
	\caption{\small The distributions of light yield across for a 10-mm-thick strip for “top” (blue) and “bottom” (green) directions (a). Normalized distribution (b) for summary of “top” and “bottom” distributions, a blue rectangular illustrates the strips geometry.}
	\label{fig: scanres10mm}
\end{figure}

The distributions for 7x40 mm strip are presented on Figure \ref{fig: scanres07mm}, and the distributions for 10x40 mm strip are presented on Figure \ref{fig: scanres10mm}. A relative light loss near the WLS fiber for the “top” position of the fiber at the middle of the strip is appeared as expected. In the opposite direction, when WLS fiber is placed far from the source, the distribution for light is smooth at the middle of the strip since most light collected at 5 mm depth. Combining “top” and “bottom” distributions, we can mimic the light yield collected by cosmic muons. 
The expected $\beta-$particles beam divergence (Fig. \ref{fig: colbeamsim}) causes some uncertainties on the transverse distribution near the strips edges.

One can see that the obtained distributions by transverse scan using collimated $\beta-$source have a close shape as for the similar distributions but obtained on cosmic rays.

%% file: parts/SimulationModule.tex
Now, having the light yield distribution function, it is possible to calculate the charged particles registration probability for CRV module using GEANT-4. In this part chapter we will calculate the registration probability for two types of CRV modules: one consists of 4 layers of 15 strips each, and the second consists of 4 layers of 4 strips each. Calculations were done for the models with 7x40 mm2 and 10x40 mm2 strips in cross-section. The scintillation light yield on cosmic muons at 2500 mm distance from SiPM for 7x40x3200 $mm^3$ and 10x40x3200 $mm^3$ strips was measured in advance and amounted to 21 ph.e. and 30 ph.e. respectively on average. These values were used in simulation to create the map of the charged particles registration probability.

\subsection{Calculation for charged particle registration probability for 4 layers by 15 strips on each CRV module}

\begin{figure}[!htbp]
	\centering
	\subfloat[]{\includegraphics[height=0.15\textheight, keepaspectratio]
		{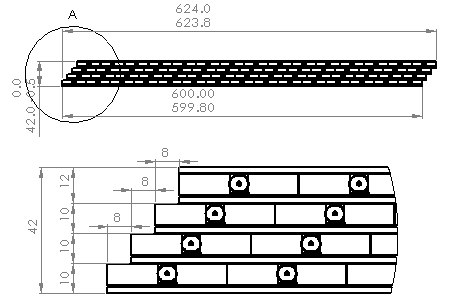} \label{fig: modulelaypat}}
	\subfloat[]{\includegraphics[height=0.15\textheight, keepaspectratio]
		{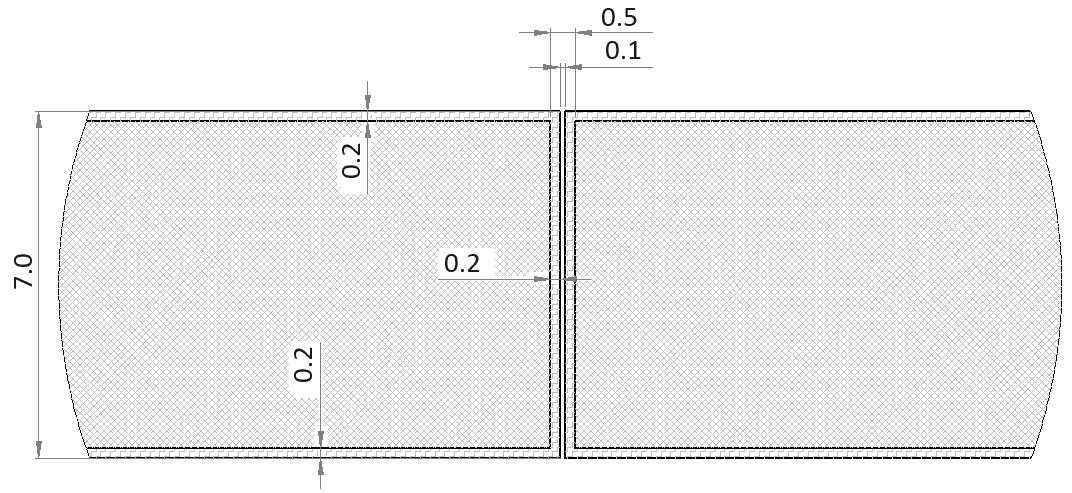} \label{fig: modulelaygap}}
	\caption{\small Layout of the CRV Module and "8-8-8" shift pattern example (a). Gaps between strips (b)}
	\label{fig: modulelay}
\end{figure}

The GIANT model for the CRV module discussed in this chapter consists of 4 layers with 15 strips. The 2-mm-thick Al sheets are placed between layers and 0.5-mm-thick gap between sheet and strips was included also – so, the 3 mm gap (0.5 mm+2mm+0.5mm) is between layers in total (Fig. \ref{fig: modulelaypat}). 

Also, the model includes gaps between scintillation volumes (so-called working volumes) for neighbor strips. This gap should include a physical gap between neighbor strips and should also include the non-scintillation space of strip created by a reflective cover for the strip. The thickness of the strip reflective cover discussed in this article is 0.2 mm. Thus, the total gap between a neighbor strip’s scintillation working volumes is 0.5 mm (0.2mm+0.1mm+0.2mm) (see \ref{fig: modulelaygap}). Each layer is shifted to another by the step and the set of such steps related to all layers creates a so-called shift pattern for CRV module geometry (Fig. \ref{fig: modulelaypat}).

The charged particles registration probability with different shift pattern of layer were studied for particles passing CRV module in various angle. Please note, that this calculation does not include the real angular distribution of the cosmic muons, or, in other words, remaining flat. But the cosmic muons angular intensities distribution can be combined later with CRV module registration probability depending on the module installation direction (horizontal, vertical, etc.) to get real estimation of the overall module efficiency to detect cosmic muons. The light yield calculated for the muon path according to the distribution we found in previous chapter by transversal scan of the strips.

\begin{figure}[!htbp]
	\begin{center}
		\includegraphics[width=0.75\textwidth, keepaspectratio]
		{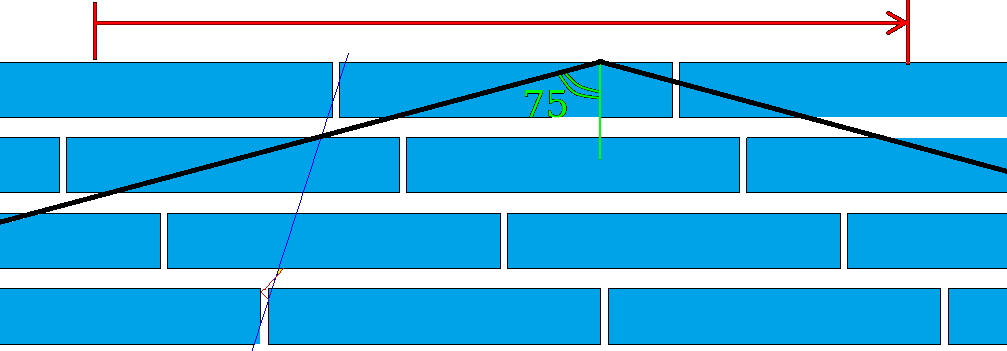}
		\caption{\small  Layout to create the registration probability map for CRV module}
		\label{fig: simlay}
	\end{center}
\end{figure}

To calculate charged particle registration probability on cosmic muon for CRV module, we created so called registration probability map, corresponding for the different muon path by angle and by beginning point of the penetration (Fig. \ref{fig: simlay}). We set the “0” on the middle of the 8th strip in top layer (the middle of the layer in a cross section). And angle “0” is the vertical to the strips. An area of “-40 to +40” mm (shown by red line on a figure) mm were studied with step of 0.1 mm. In each step, the angles were varied within a cone from -75 to +75 degree with 1 degree step (orange lines on Figure \ref{fig: simlay} represent the borders for the cone). The simulations of the muon registration probability for the CRV module with different pattern of shifts layers to each other were done. The average light yield set to 21 ph.e. and the threshold (level of discrimination) set to 5 ph.e.

\begin{figure}[!htb]  
 	\centering
 	\subfloat["20-20-20"]{\includegraphics[height=0.15\textheight, keepaspectratio]
 		{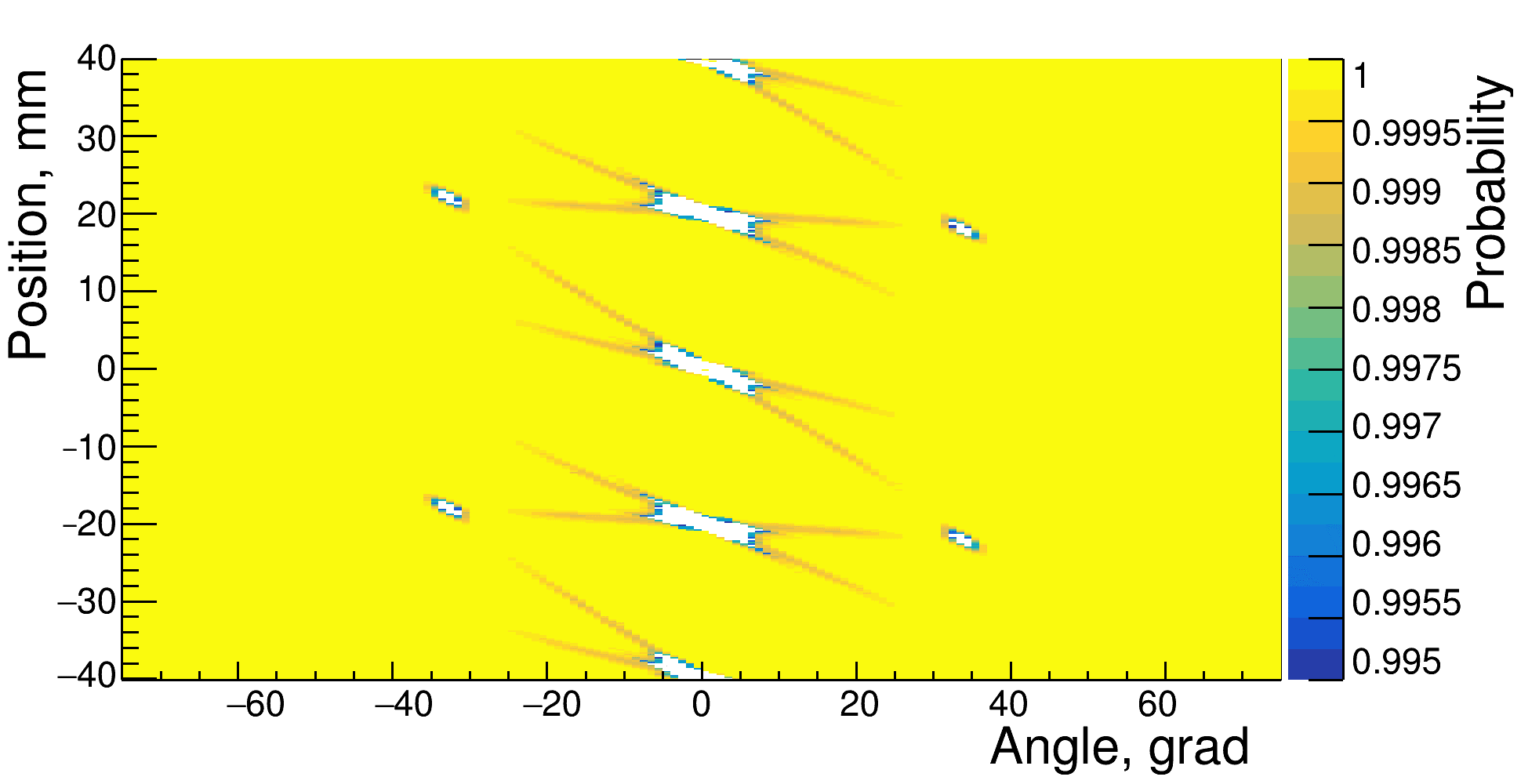}}
 	\hfill
 	\subfloat["10-10-10"]{\includegraphics[height=0.15\textheight, keepaspectratio]
 		{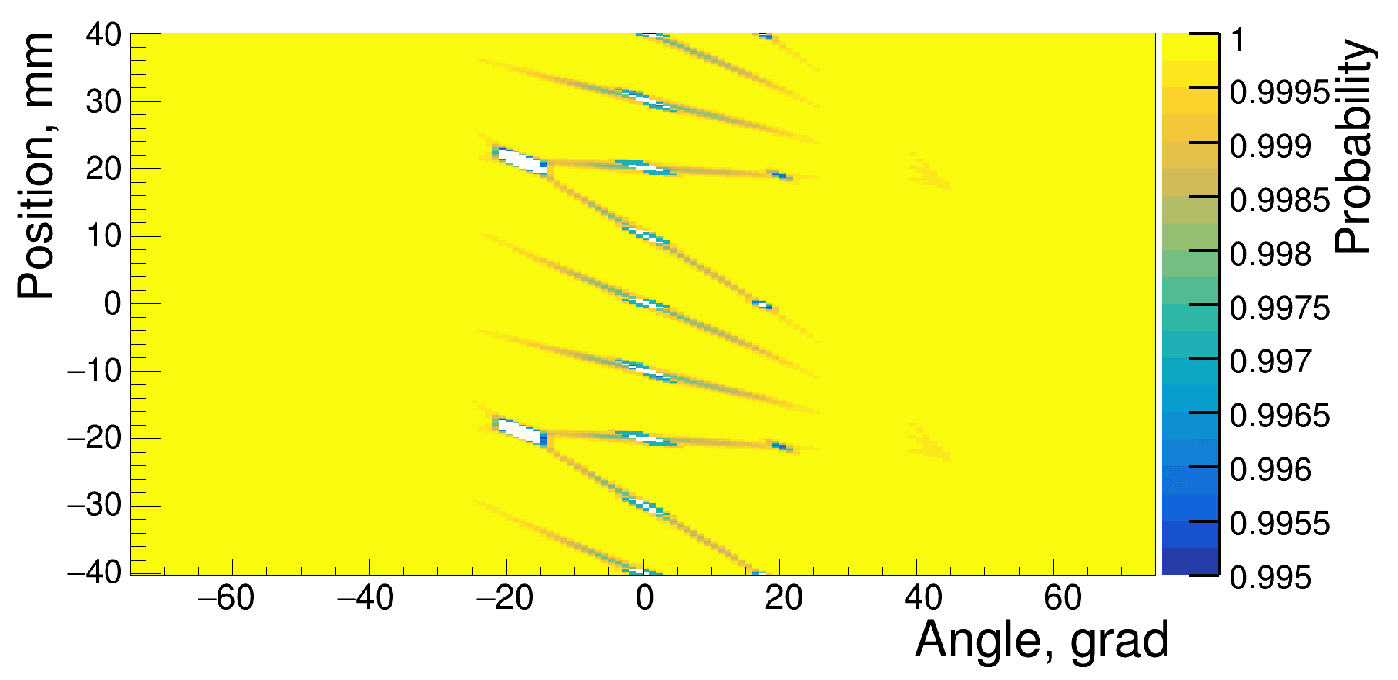}}
 	\hfill
 	\subfloat["8-8-8"]{\includegraphics[height=0.15\textheight, keepaspectratio]
 		{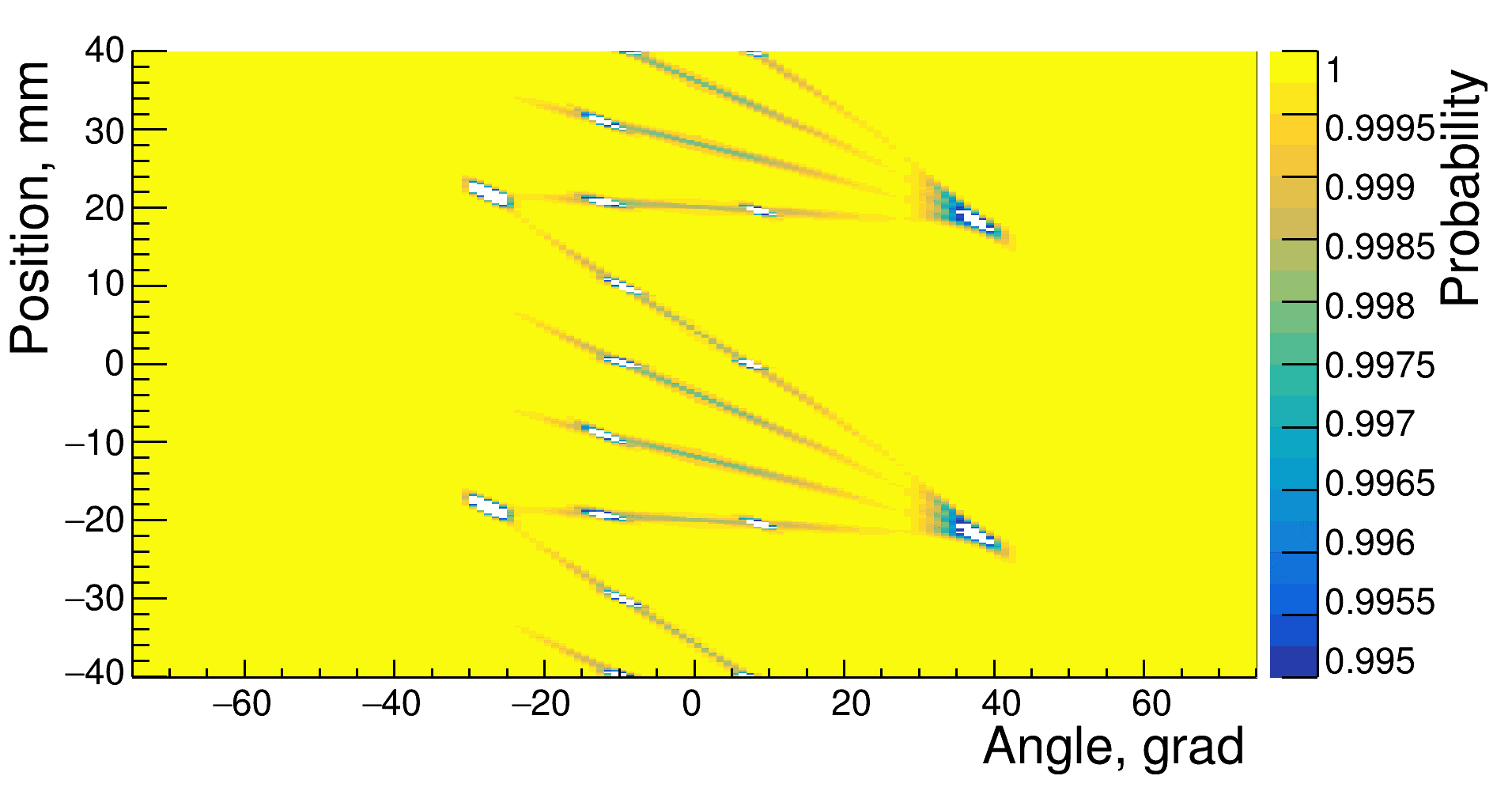}}
 	\hfill
 	\subfloat["8-10-8"]{\includegraphics[height=0.15\textheight, keepaspectratio]
 		{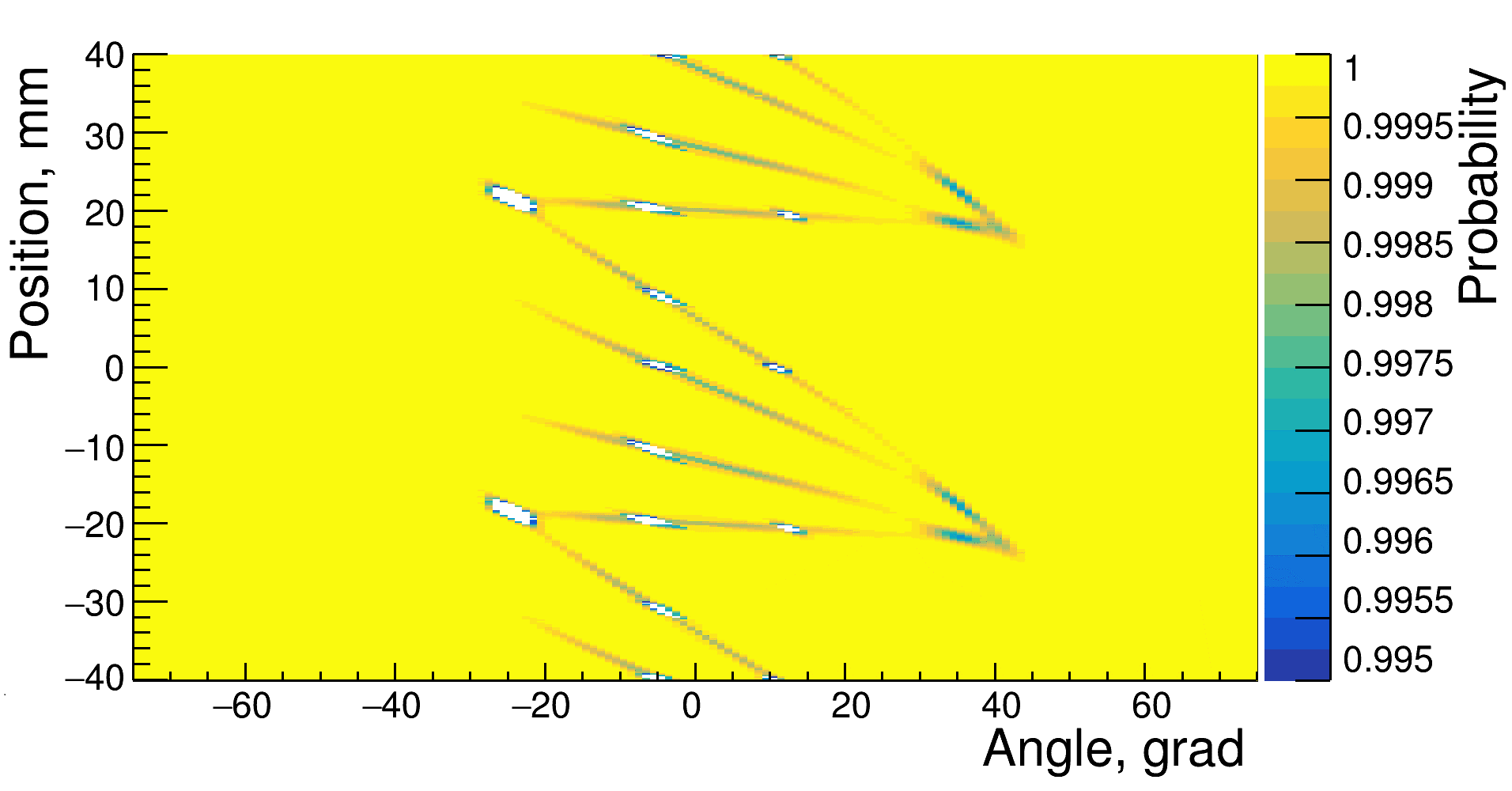}}
 	\hfill
 	\subfloat["8-10-10"]{\includegraphics[height=0.15\textheight, keepaspectratio]
 		{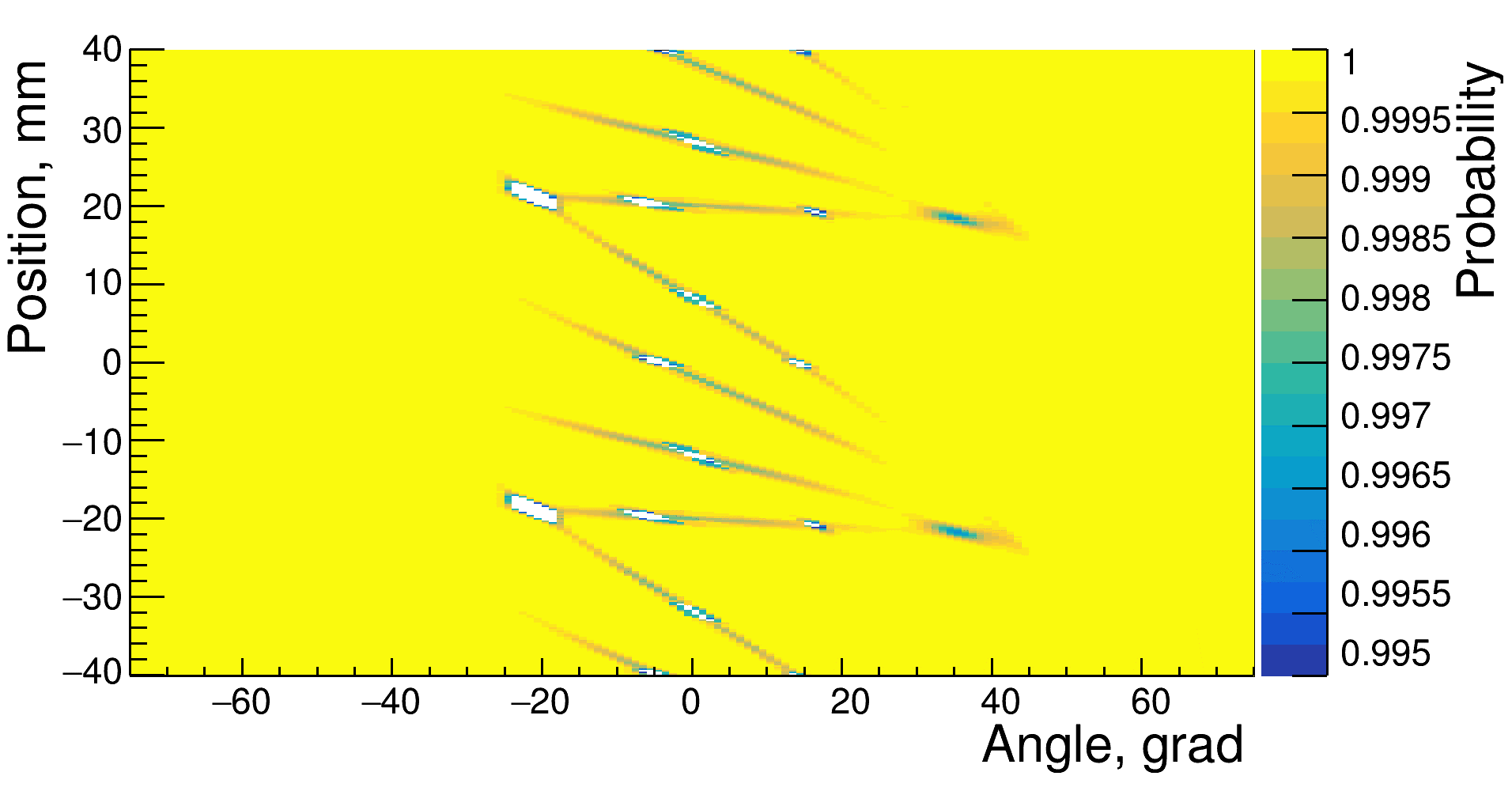}}
 	\hfill
 	\subfloat["8-10-18"]{\includegraphics[height=0.15\textheight, keepaspectratio]
 		{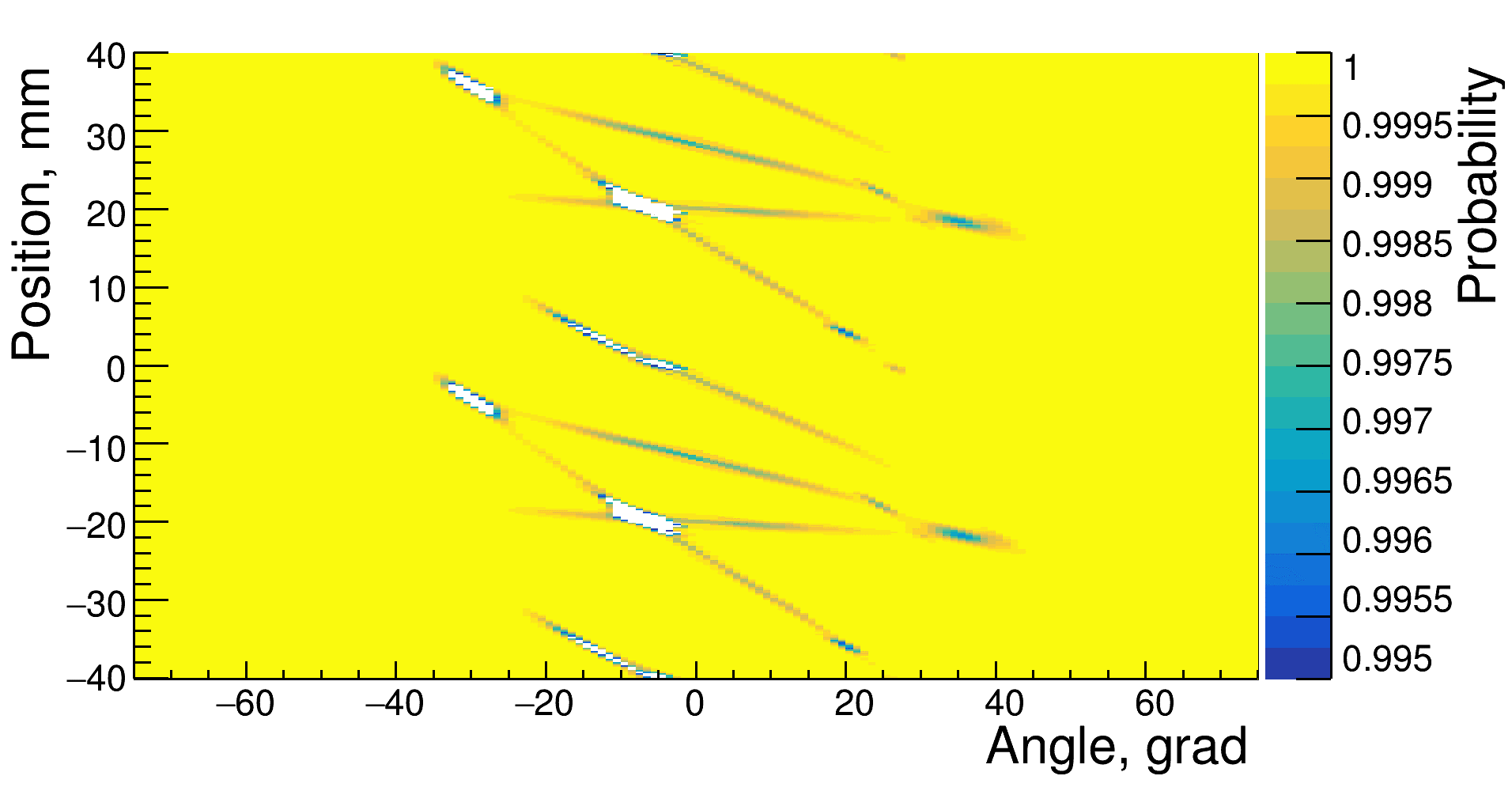}}
 	\caption{\small Examples of registration probability maps for various CRV modules shift patterns: an original "20-20-20" shift pattern by TDR2020 (a)}
 	\label{fig: regprobmap}
 \end{figure}

The registration probability maps (2D distribution by the angle and by point of entrance) were created for more than 60000 patterns of CRV modules, and, overall, calculations took about 1 month. The examples of registration probability maps for various shift patterns for CRV modules based on 7x40 $mm^2$ in cross-section strips were illustrated on Figure \ref{fig: regprobmap}. White areas on plots represent the areas with registration probability less than 99.5\%. One can see that it is difficult to achieve 99.99\% efficiency for the module with current configuration of the 7-mm-thick strip.

Table \ref{table: shiftpat} presents the overall efficiency with no angular distribution for cosmic muons included for the best shift patterns with the average light yield set to 21 ph.e. and the threshold (level of discrimination) set to 5 ph.e.

\begin{table}[!htb]
\fontsize{8pt}{8pt}\selectfont
\centering
    \begin{tabular}{|l|l|}
    \hline
    Shift   pattern & Overall   efficiency \\ \hline
    20-20-20   mm   & 0.99902              \\ \hline
    10-10-10   mm   & 0.99976              \\ \hline
    8-8-8   mm      & 0.99988              \\ \hline
    8-10-6   mm     & 0.99983              \\ \hline
    8-10-8   mm     & 0.99985              \\ \hline
    8-10-10   mm    & 0.99981              \\ \hline
    8-10-12   mm    & 0.99976              \\ \hline
    8-10-14   mm    & 0.99968              \\ \hline
    8-10-16   mm    & 0.99963              \\ \hline
    8-10-18   mm    & 0.99958              \\ \hline
    \end{tabular}
    \caption{\small Best shift patterns for CRV modules based on 7x40 $mm^2$ strips. The average light yield set from the strip to 21 ph.e.  and the threshold set to 5 ph.e.}
    \label{table: shiftpat}
\end{table}

\begin{figure}[!htbp]
	\centering
	\subfloat[]{\includegraphics[height=0.15\textheight, keepaspectratio]
		{figures/png/shp-8-8-8_mm.png} \label{fig: regcomp07}}
	\subfloat[]{\includegraphics[height=0.15\textheight, keepaspectratio]
		{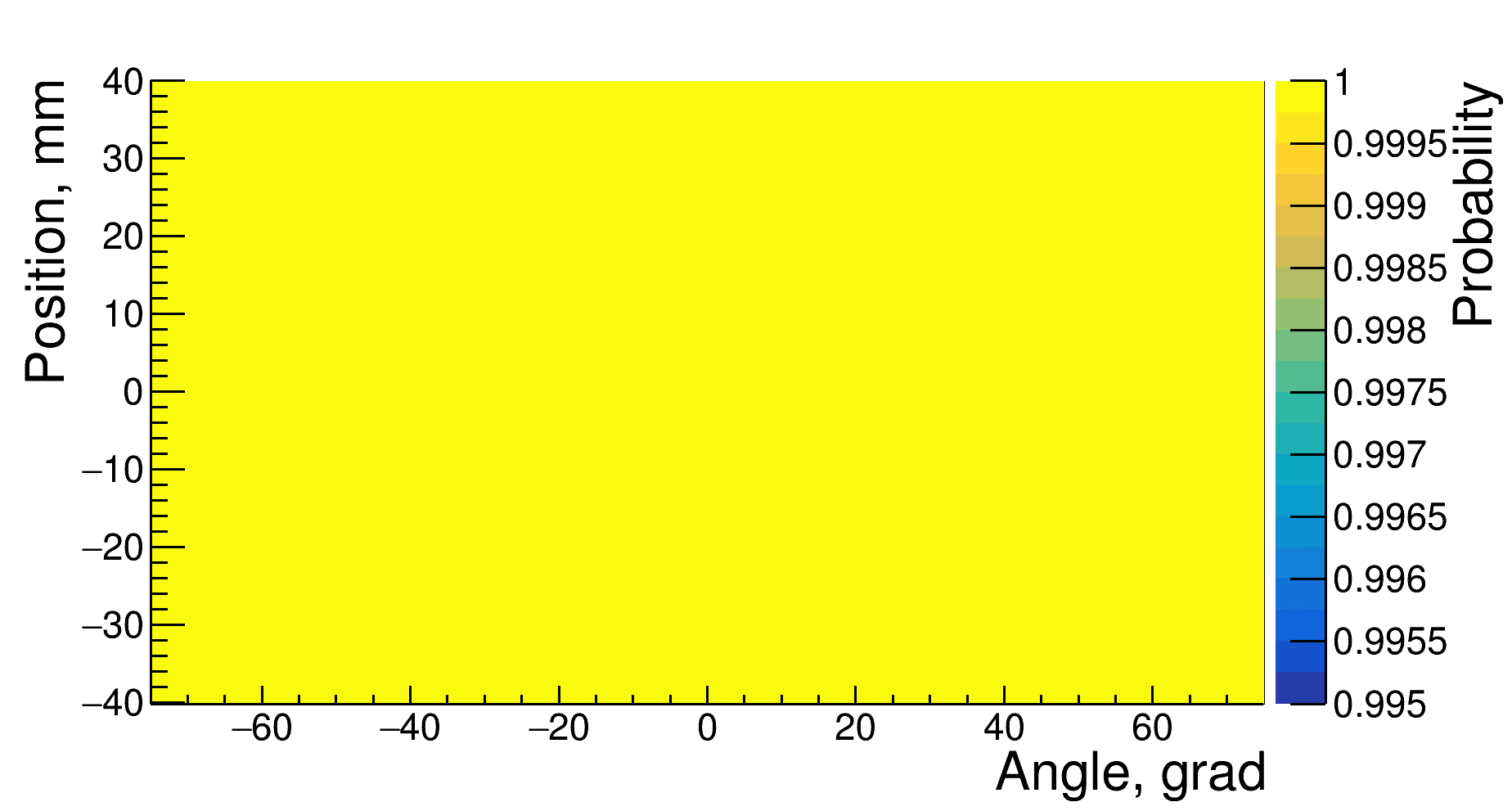} \label{fig: regcomp10}}
	\caption{\small The registration probability maps for CRV modules with 7-mm-thick (a) and on 10-mm-thick (b) strips. Strips width and shift pattern are same for both cases: 40 mm and “8-8-8” respectively.}
	\label{fig: regcomp}
\end{figure}

We also did similar simulation for CRV modules based on 10x40 $mm^2$ in cross-section strips. In this case, gaps, way to change particle propagation, etc. left same, but the average light yield set to 30 ph.e. but the threshold (level of discrimination) remained same: 5 ph.e.. The results for 10-mm-thick case shows that the charged particles registration probability was better than 99.99\%. The comparison of the plots for 7-mm-thick and 10-mm-thick strips is shown in Fig. \ref{fig: regcomp}. 

One can see that the 10-mm-thick strip has almost 99.999\% overall registration probability in comparison to 7-mm-thick strip, which has 99.988\%. Two major factors help to achieve such a difference. On one side, the 10-mm-thick strip has better  strip thickness to gap between strips ratio under the same conditions – 20 (10/0.5) against 14 (7/0.5) for 7-mm-thick strip. Other also very important factor is that 10-mm-thick strip has about 30\% more light yield against to 7-mm-thick strip: 30 ph.e. vs 21 ph.e..

\subsection{Influence of the natural aging on charged particle registration probability for CRV module}

As we know, the plastic scintillation counter experiences the deterioration of the light yield by the time even in case it never used in experiments, or so-called natural aging. The speed of the deterioration (or how much loss in percent per year) is very important and should be considered when such a system should be used for an extended period, or the detector will be used after some notable delay (for instance, years) once it was created.

Our 12-year-long study \cite{scintCDFaging} for various geometry of scintillation counters (some of it equipped with WLS fibers as well) used in CDF experiment shows that the light yield decreases with a rate of 6…9\% per year for polystyrene (PS) based scintillation counters. A major result taken from this study is that the natural deterioration of just PS without other factors is about 6.6\% per year. So, we can expect, that the PS-based strips deterioration will be on same, 6…7\% per year, level in best case.

We should note that the results of the registration probability presented above are based on the 21 ph.e. and 30 ph.e. light yield for 7-mm-strip and 10-mm-strip. But the strips experience the natural aging process, or the light yield deteriorates with time. This fact shows that it is necessary to consider the natural aging of the scintillation strips to predict the efficient lifetime of the strips working under the requested conditions by registration efficiency, or to determine their long-term stability. 

\begin{figure}[!htb]  
 	\centering
 	\subfloat["3-year term for 7-mm-thick strip"]{\includegraphics[height=0.15\textheight, keepaspectratio]
 		{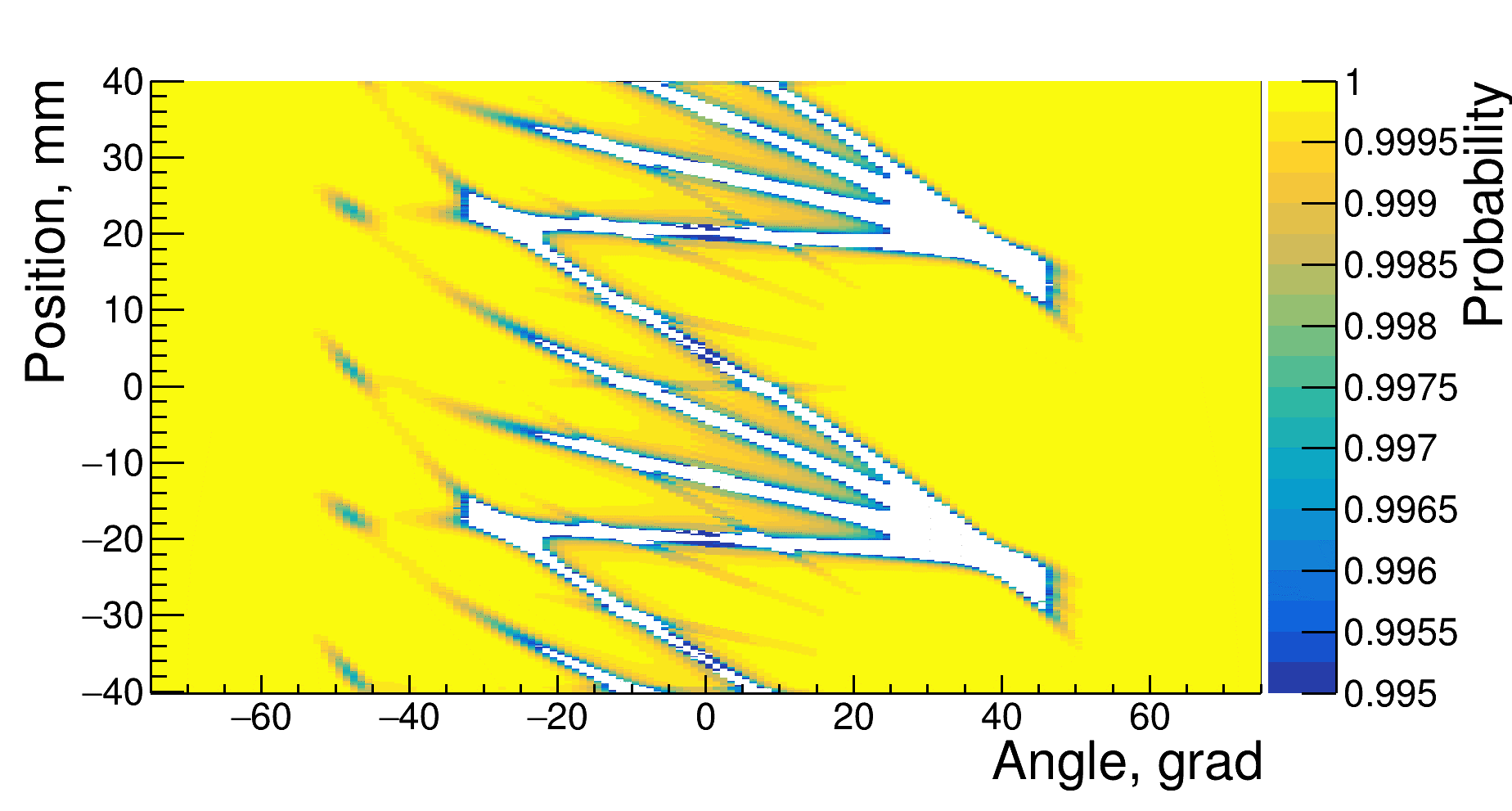}}
 	\hfill
 	\subfloat["3-year term for 10-mm-thick strip"]{\includegraphics[height=0.15\textheight, keepaspectratio]
 		{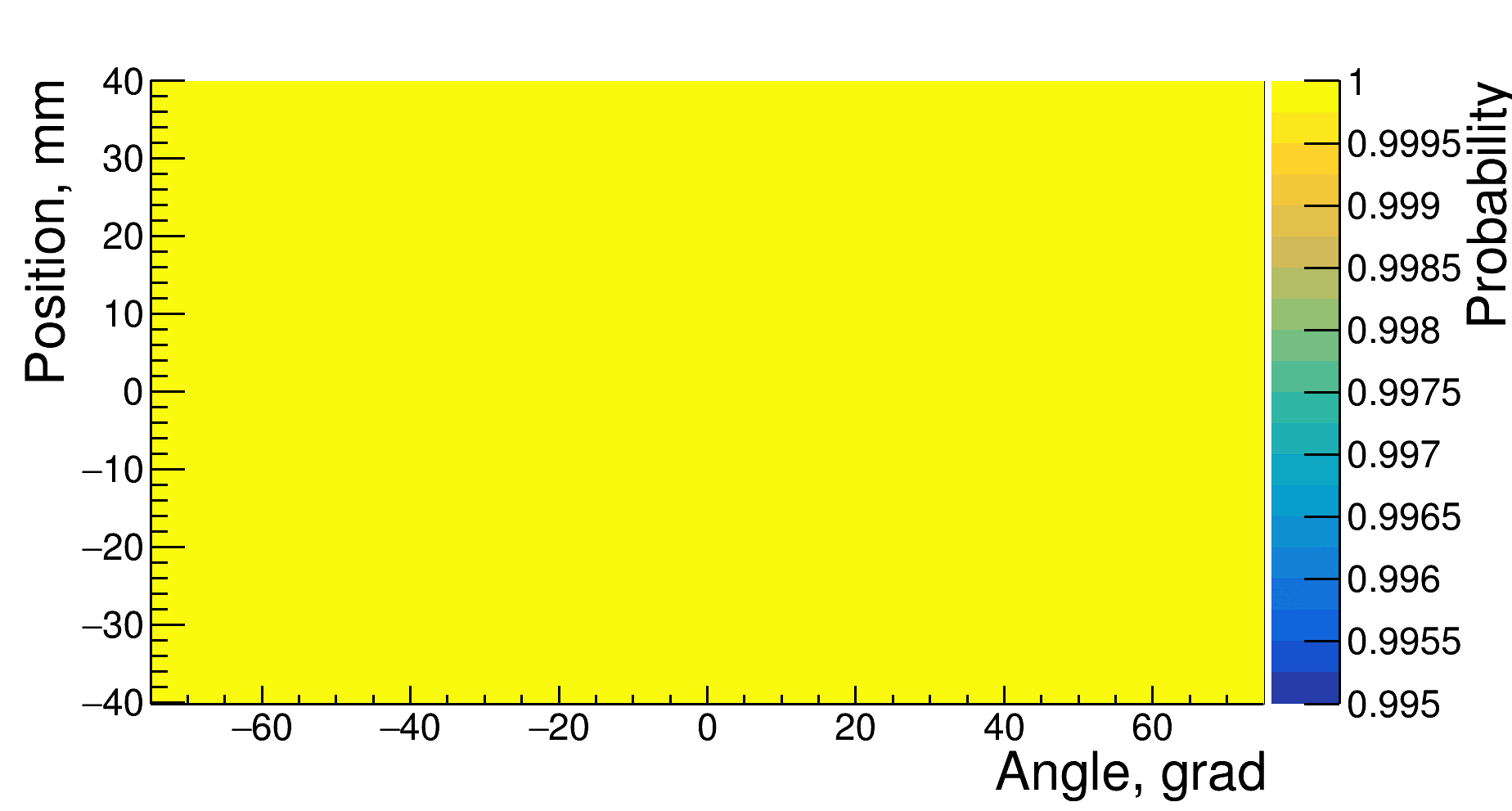}}
 	\hfill
 	\subfloat["13-year term for 7-mm-thick strip"]{\includegraphics[height=0.15\textheight, keepaspectratio]
 		{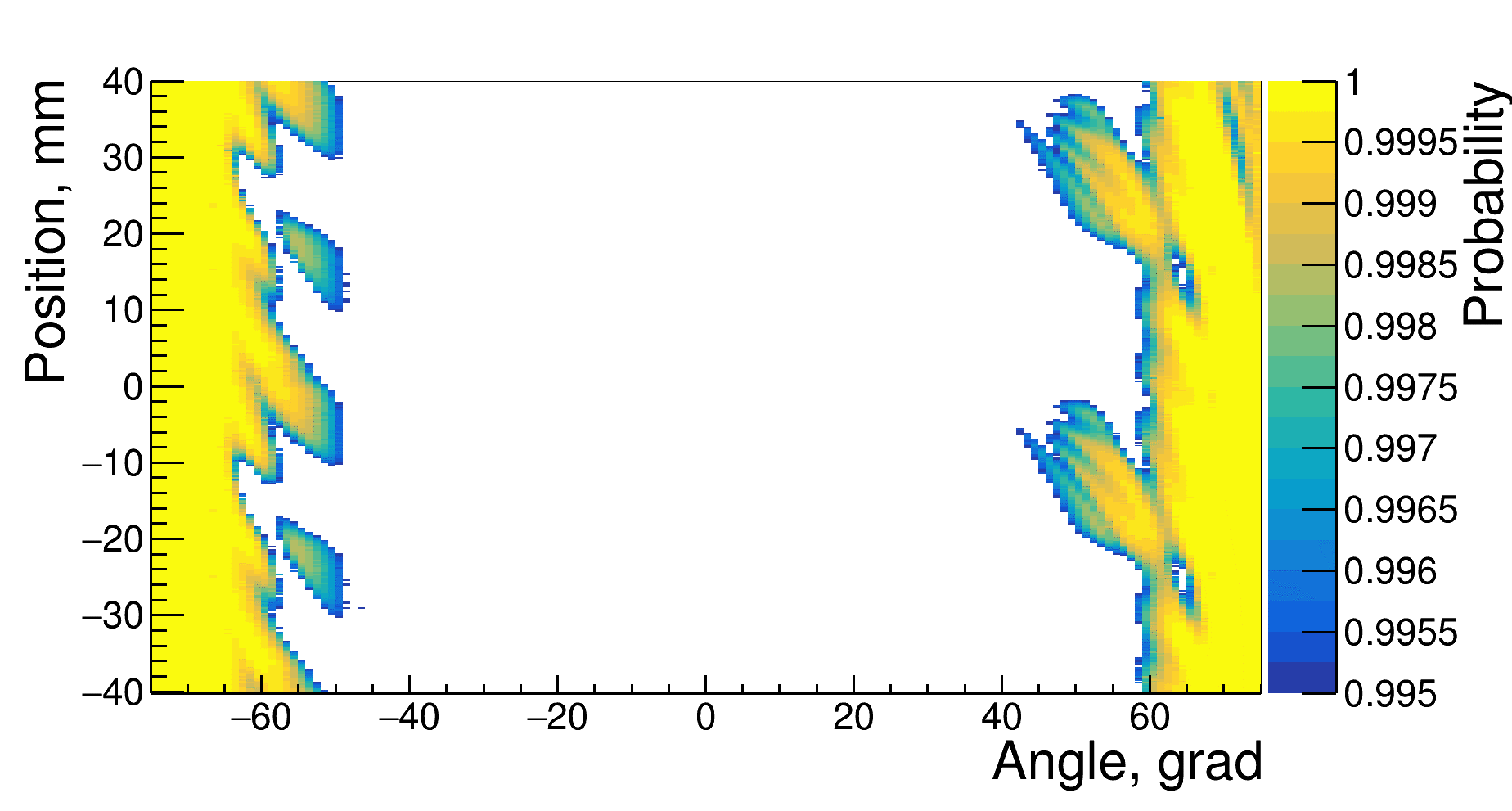}}
 	\hfill
 	\subfloat["13-year term for 13-mm-thick strip"]{\includegraphics[height=0.15\textheight, keepaspectratio]
 		{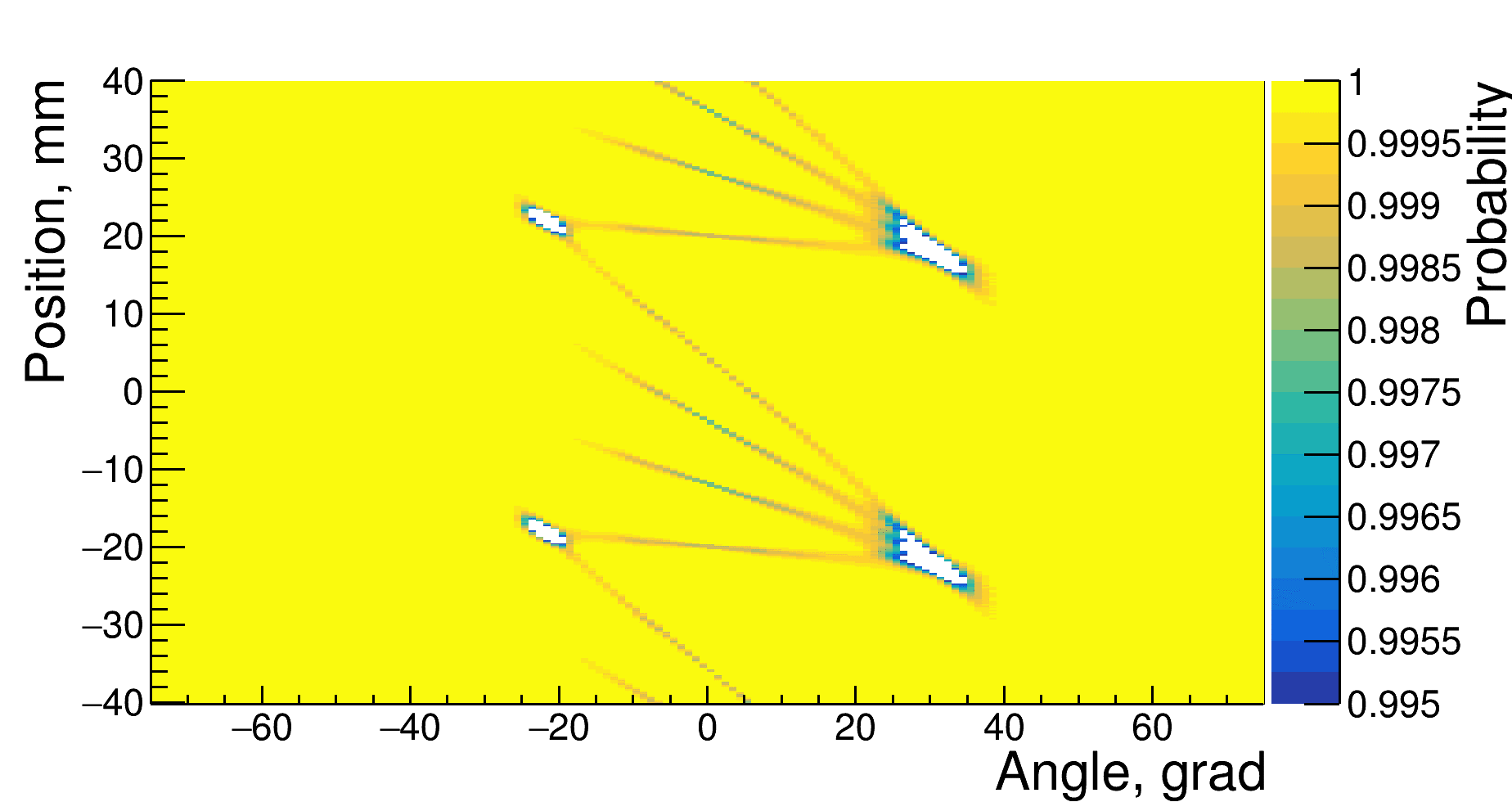}}
 	\hfill
	\caption{\small The registration probability maps for CRV modules with 7-mm-thick (a) and on 10-mm-thick (b) strips after 3-year natural aging and plots (c) and (d) for 13-year natural aging. Other conditions are the same.}
	\label{fig: regcompaging}
\end{figure}

If we consider the strip aging rate just as 6\% per year, the light yield for 7- and 10-mm-thick strips will drop respectively to 17.6 ph.e. and 25.2 ph.e. for a 3-year term, and to 9.8 ph.e. and 14.1 ph.e. for a 13-year term. The registration probability maps for 3-year and 13-year terms are shown on Figure \ref{fig: regcompaging} and overall efficiencies of the module are presented in Table \ref{table: effaging}.

\begin{table}[!htb]
\fontsize{8pt}{8pt}\selectfont
    \begin{tabular}{|l|l|l|}
    \hline
    Pattern: 8-8-8                                      & 7-mm strips case & 10-mm strips   case \\ \hline
    Initial light   yield                               & 21 ph.e.         & 30 ph.e.            \\ \hline
    Overall   efficiency, initial state                 & 99.925 \%        & 100.000 \%          \\ \hline
    Overall   efficiency, 3-year-aging by 6\% per year  & 99.670 \%        & 99.999 \%           \\ \hline
    Overall   efficiency, 13-year-aging by 6\% per year & 83.166 \%        & 99.984 \%           \\ \hline
    \end{tabular}
    \caption{\small Expected overall efficiency for CRV modules considering the 6\%-per-year natural aging (no angular distribution for cosmic muons included)}
    \label{table: effaging}
\end{table}

%% file: parts/ExpStudy4x4.tex
To perform this study, we prepared 16 3.2-m-long scintillation strips with 7x40 $mm^2$ in cross-section, and 1.2-mm diameter Kuraray Y11 (200) WLS fiber glued into the central groove. Strips were produced by Uniplast also. So, as a next step, we created from scratch the 4x4 CRV module prototype using these strips: 4 layers with 4 strips on each sliced with steel 2-mm thick sheets (Fig.\ref{fig:  crv4x4layout}) and tried to compare the efficiency of this 4x4 CRV module obtained by exposition on cosmic rays with simulation result obtained using GEANT-4.

\begin{figure}[!htbp]
	\centering
	\subfloat[]{\includegraphics[height=0.14\textheight, keepaspectratio]
		{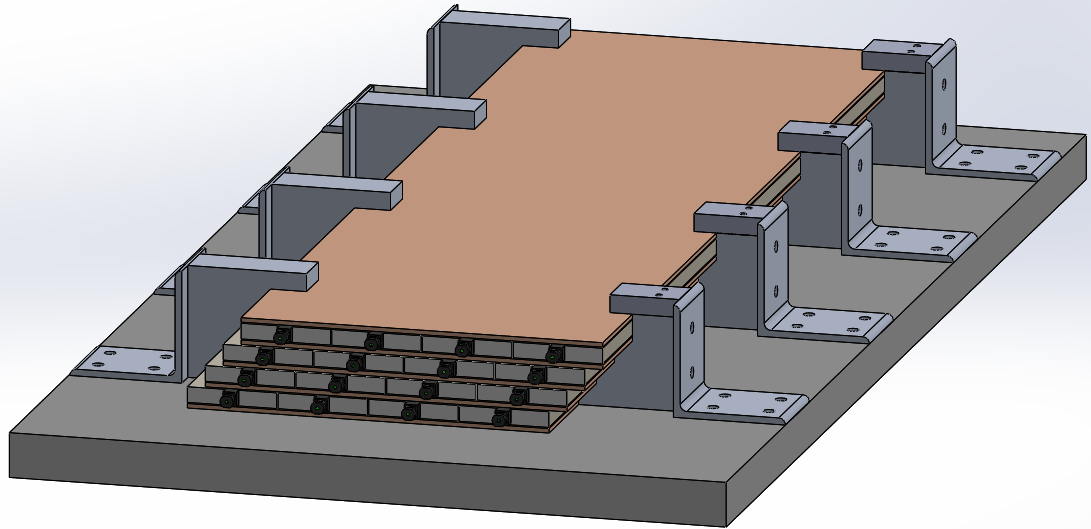}}
	\hfill
	\subfloat[]{\includegraphics[height=0.14\textheight, keepaspectratio]
		{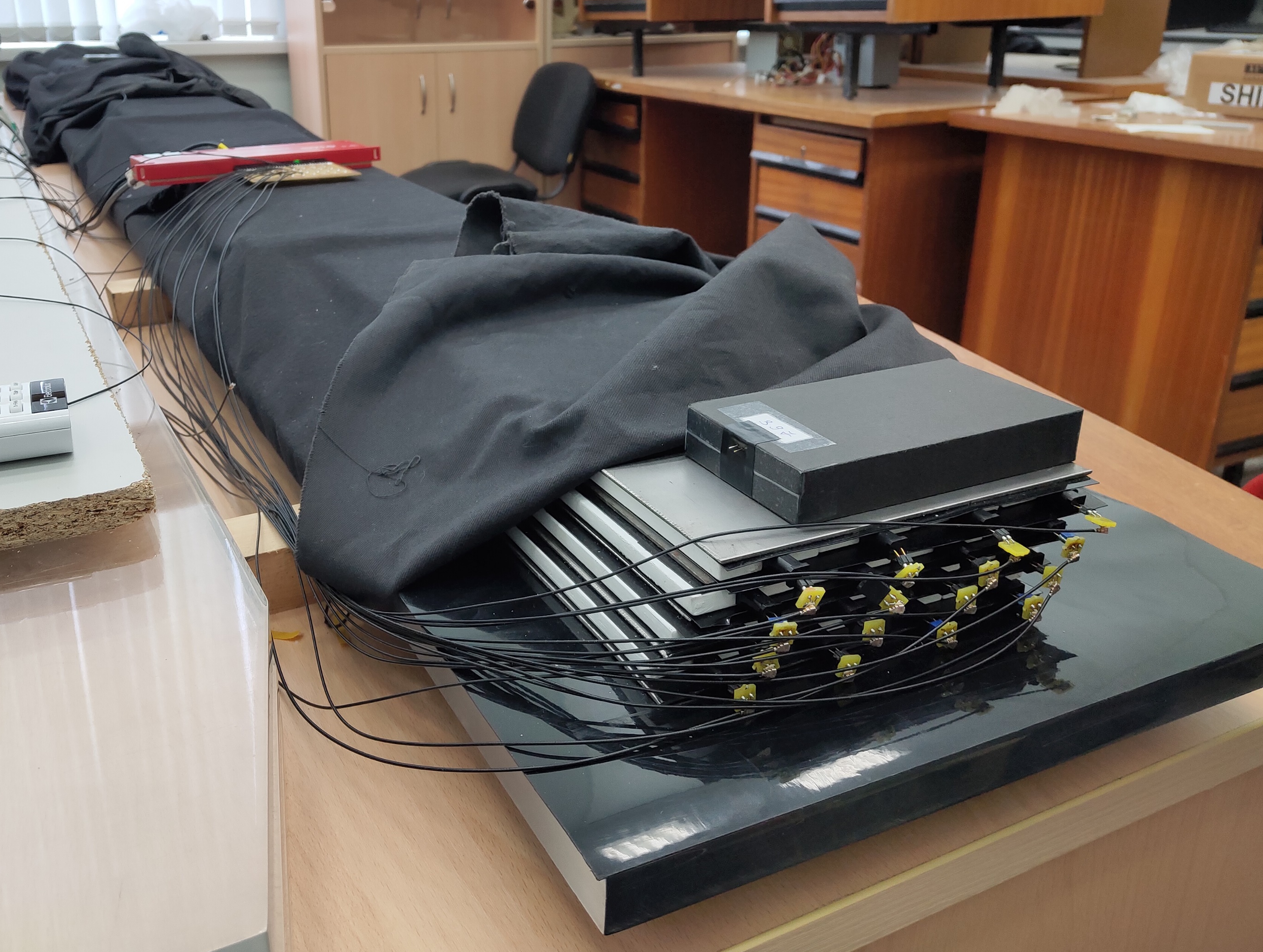}}
	\caption{\small Layout of 4x4 CRV module prototype (a) and real look of it (b)}
	\label{fig: crv4x4layout}
\end{figure}

\subsection{Testing the geometry of the strips}

Before creation of the 4x4 CRV module prototype, it is necessary to check the strips geometrical parameters. These real values were needed to put proper values into the simulation as model geometrical parameters for CRV module.  

\begin{figure}[!htbp]
	\centering
    \includegraphics[width=0.75\linewidth, keepaspectratio]
		{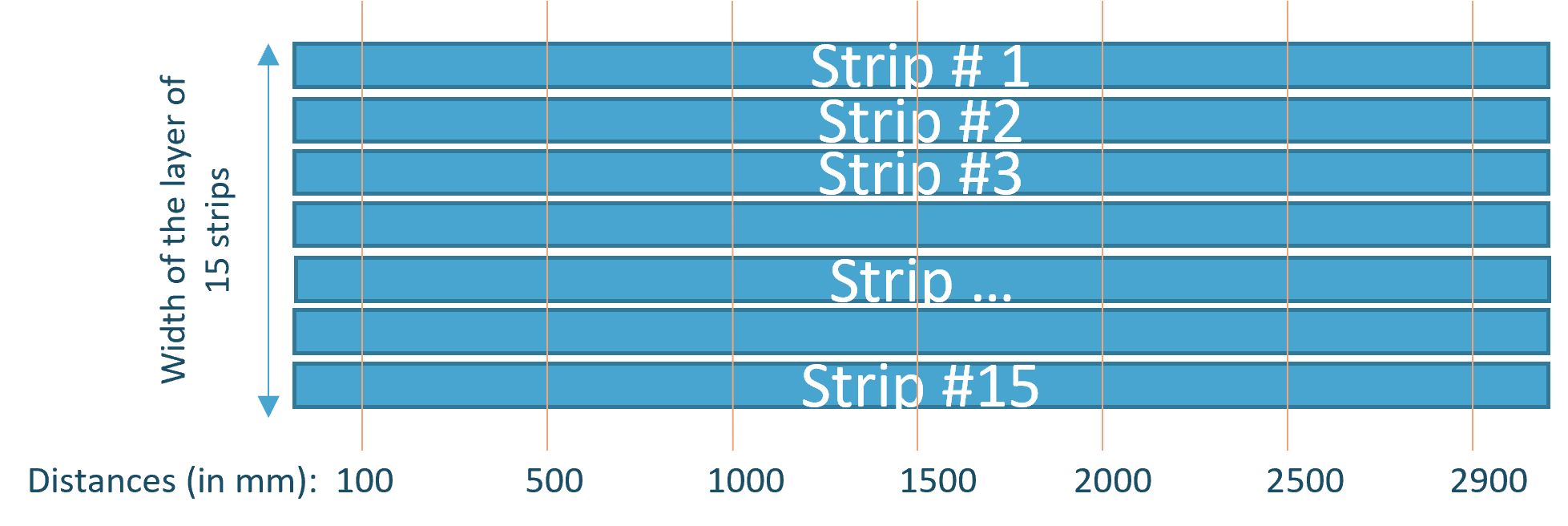}
	\caption{\small Illustration of the positions to measure widths and gaps for 4x4 CRV module prototype}
	\label{fig: gapmeasure}
\end{figure}

\begin{figure}[!htbp]
	\centering
	\subfloat[]{\includegraphics[height=0.20\textheight, keepaspectratio]
		{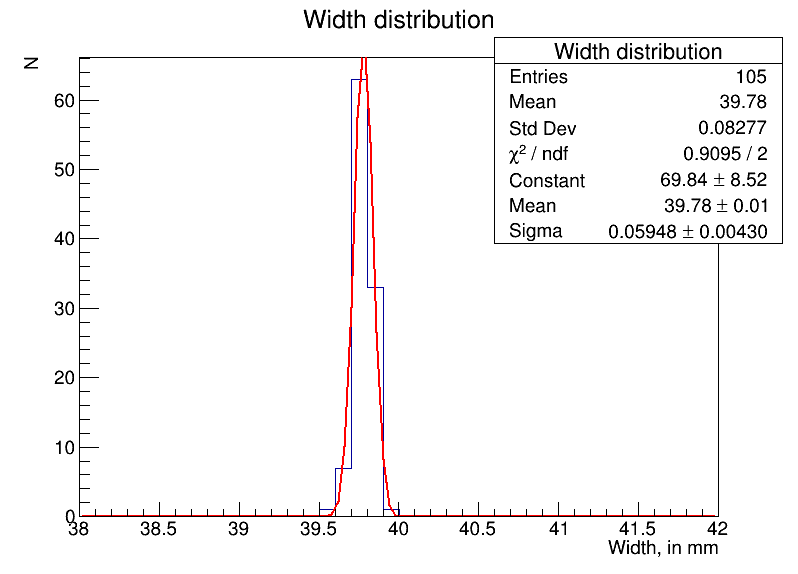} \label{fig: geoparwidth}}
	\subfloat[]{\includegraphics[height=0.20\textheight, keepaspectratio]
		{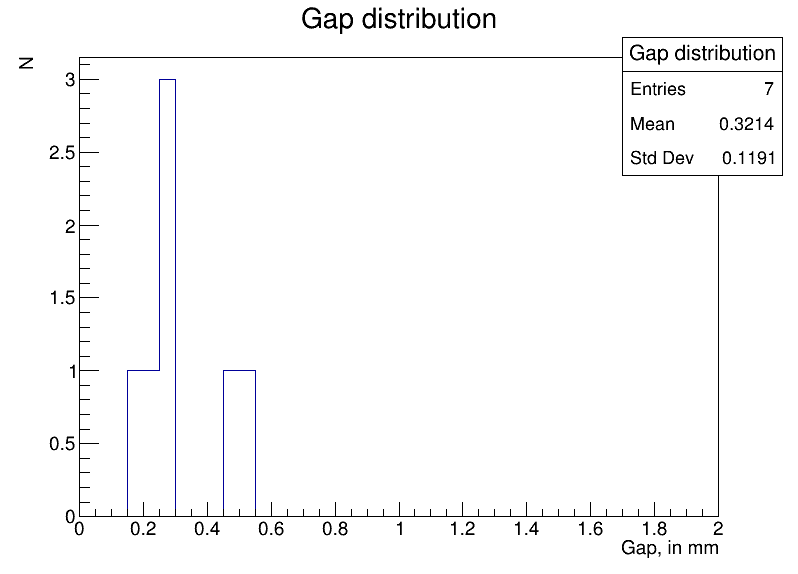} \label{fig: geopargap}}
	\caption{\small Variation of the width (a) and gaps (b) for 4x4 CRV module prototype}
	\label{fig: geopar}
\end{figure}

The width of each strip was measured at 7 positions: at distances of 100, 500, 1000, 1500, 2000, 2500 and 2900 mm from one edge of the strip (Fig. \ref{fig: gapmeasure}). The resulting distribution was approximated by Gaussian and an average value of the strip’s width was found to be equal to 39.78 mm with sigma of 0.08 mm (Fig. \ref{fig: geoparwidth}). 

Next important parameter we need to know is the real gap between strips. It was calculated as follows: we measured the width of the layer created by 15 strips on each 7 positions (Fig. \ref{fig: gapmeasure}). Then the sum of the strips widths on corresponding position which already measured was subtracted and result divided by 14 according to the number of gaps. The average gap’s value between strips was found to be about 0.32 mm $\pm$ 0.12 mm (Fig. \ref{fig: geopargap}).

\subsection{Efficiency calculation for a 4x4 CRV module on cosmic muons}

\begin{figure}[!htbp]
	\begin{center}
		\includegraphics[width=0.75\textwidth, keepaspectratio]
		{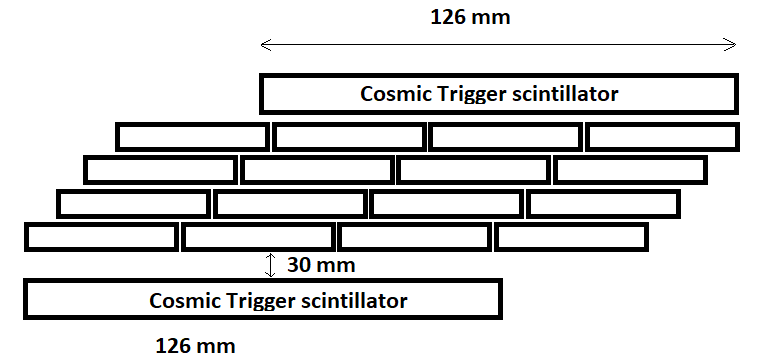}
		\caption{\small Layout of the setup for 4x4 CRV module with cosmic telescope}
		\label{fig: crvcoslay}
	\end{center}
\end{figure}

We studied 4x4 CRV modules efficiency on cosmic muons. Two 126x60x20 $mm^3$ scintillation counters create the muon telescope and these detectors were placed over and below of the module as shown at Figure \ref{fig: crvcoslay}: one telescope counter was located directly on the top of CRV module close to the module side edge; other telescope was located under the plate holding the CRV module at 30 cm distance from the bottom of the module, and close to the other side edge of the module.  The telescope was positioned at 2500 mm distance from strips readout, at far end. Such arrangement of telescope counters ensures the passage of the cosmic muon through all 4 working layers.

Light collected by SiPM (Hamamatsu S13360-1350CS; similar SiPMs were used for light collection from the strips and trigger counters). Kuraray 1.4-mm diameter optical-clear fibers were attached far from SIPM ends of the strips. Using these fibers, the flashing UV light was delivered to strips for calibrations purpose by single photon counting. Data from the module and telescope counters collected using CAEN DT5702 32-ch MPPC/SiPM readout Front-End. UV light LED was produced by HVSys light flashes calibrated LED source \cite{hvsys}. 

\begin{figure}[!htbp]
	\begin{center}
		\includegraphics[width=1\linewidth, keepaspectratio]
		{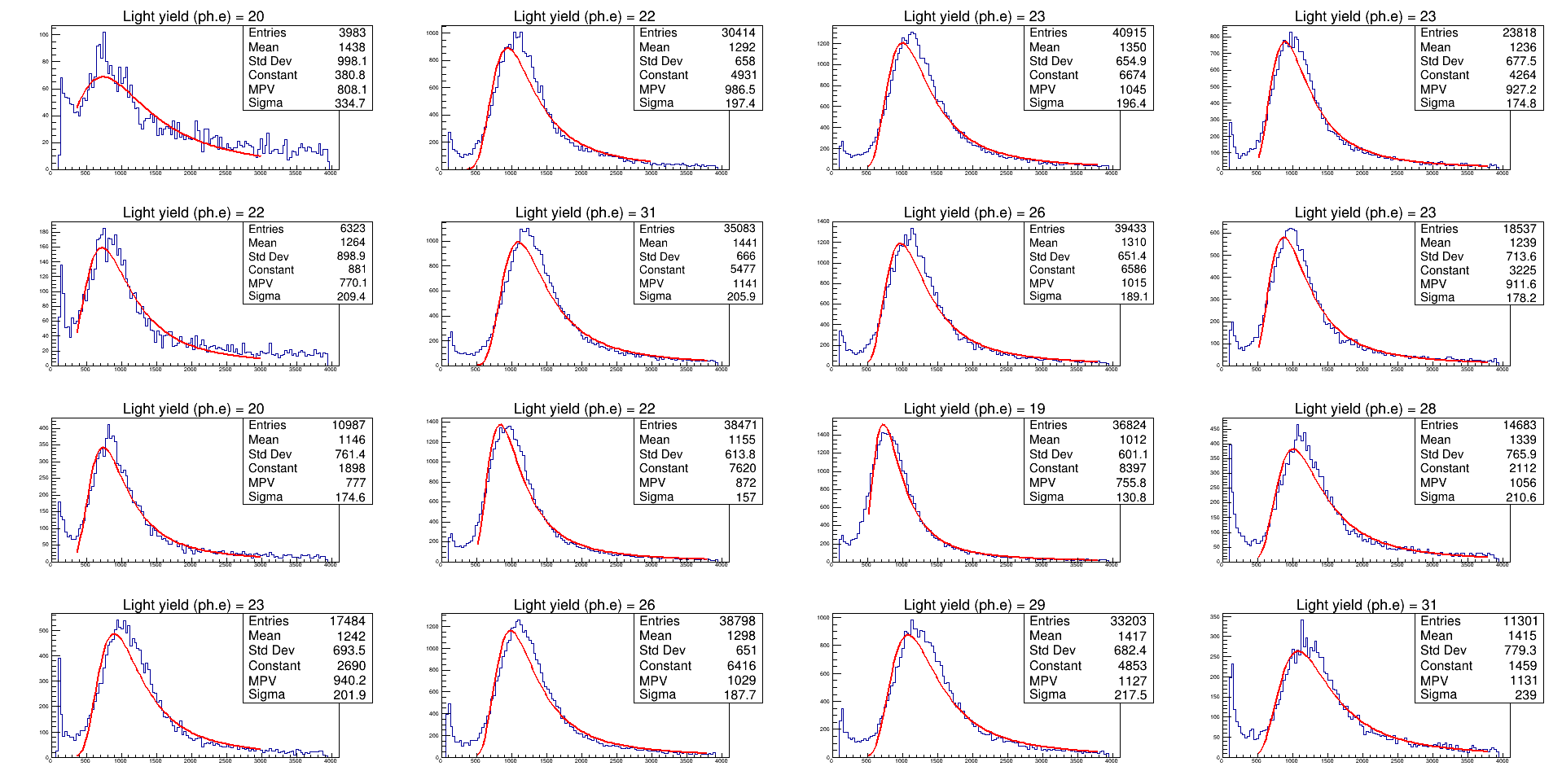}
		\caption{\small Channel by channel light yield for 4x4 CRV module}
		\label{fig: crv4x4cos}
	\end{center}
\end{figure}

Cosmic data for CRV module was taken continuously for a week duration thus allowing us to collect about 200 000 events. Using an absolute calibration method, all 16 SiPMs gain tuned for around 45 ADC channel per photon. The light yield distributions for each channel are shown in Figure \ref{fig: crv4x4cos}. After pedestal subtraction and approximation by Landau distribution, an average light yield at 250 cm distance was found as 21±3 ph.e.

Efficiency for this CRV module calculated as a ratio of the CRV module events selected by coincidence of any 3 layers of 4 to the total number of events registered by cosmic muon telescope. Data were processed offline and the threshold level set at 5 ph.e. for all channels during this analysis. Overall CRV module efficiency to register cosmic muons was found on the 99.69\% level.

\subsection{Efficiency simulation for a 4x4 CRV module}

As noted in the previous chapter, all real geometrical parameters of 4x4 CRV module were measured to set the sketch at GEANT-4 geometry properly. Also, the light yield is set to 21 ph.e. according to the data obtained from real CRV module. These parameters allowed us to provide proper simulation using Geant 4. The registration probability map for 4x4 CRV module was calculated using SLYD method as described at chapter \ref{transscan}. 

\begin{figure}[!htbp]
	\centering
	\subfloat[]{\includegraphics[height=0.15\textheight, keepaspectratio]
		{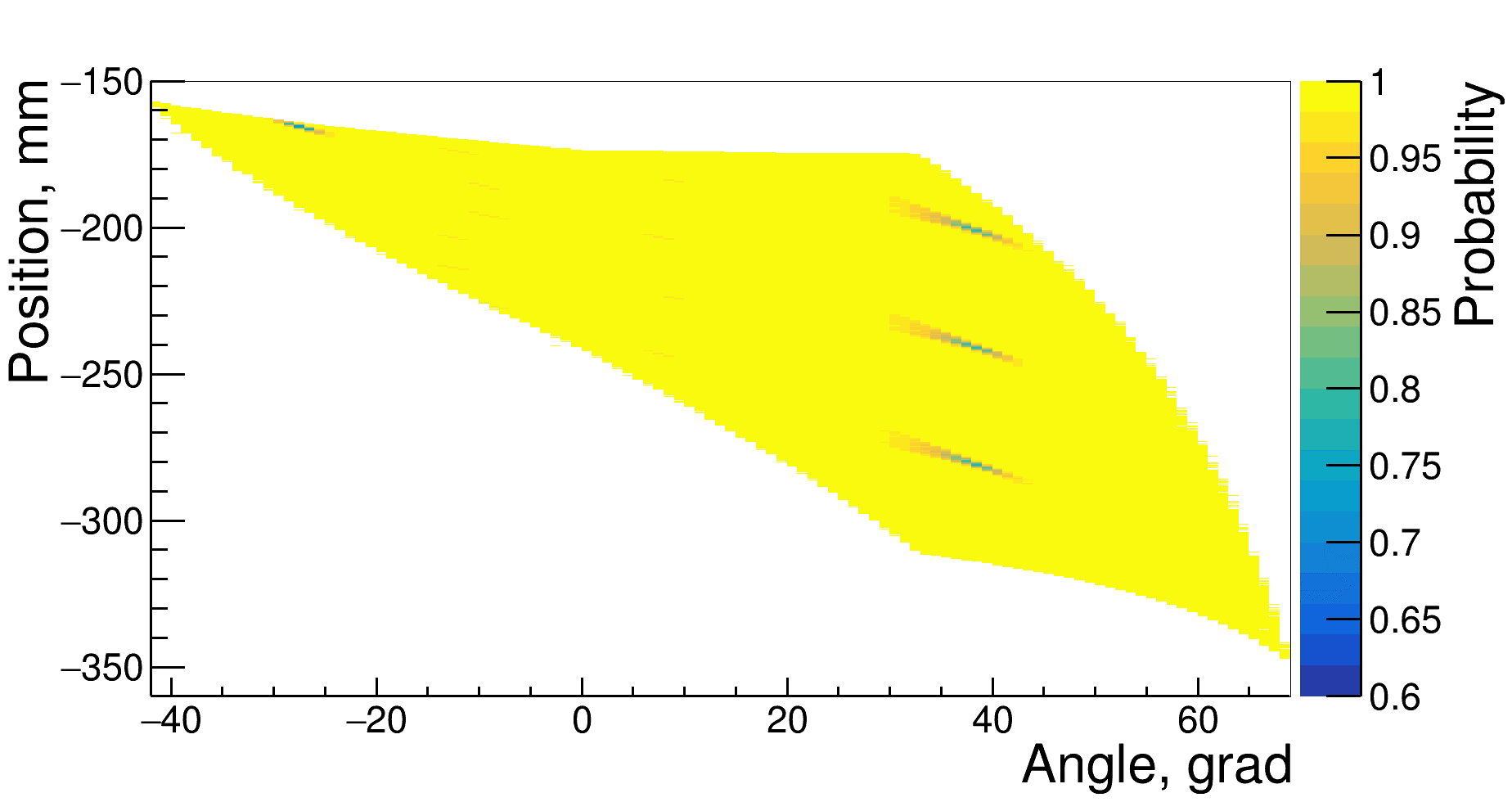}}
	\hfill
	\subfloat[]{\includegraphics[height=0.15\textheight, keepaspectratio]
		{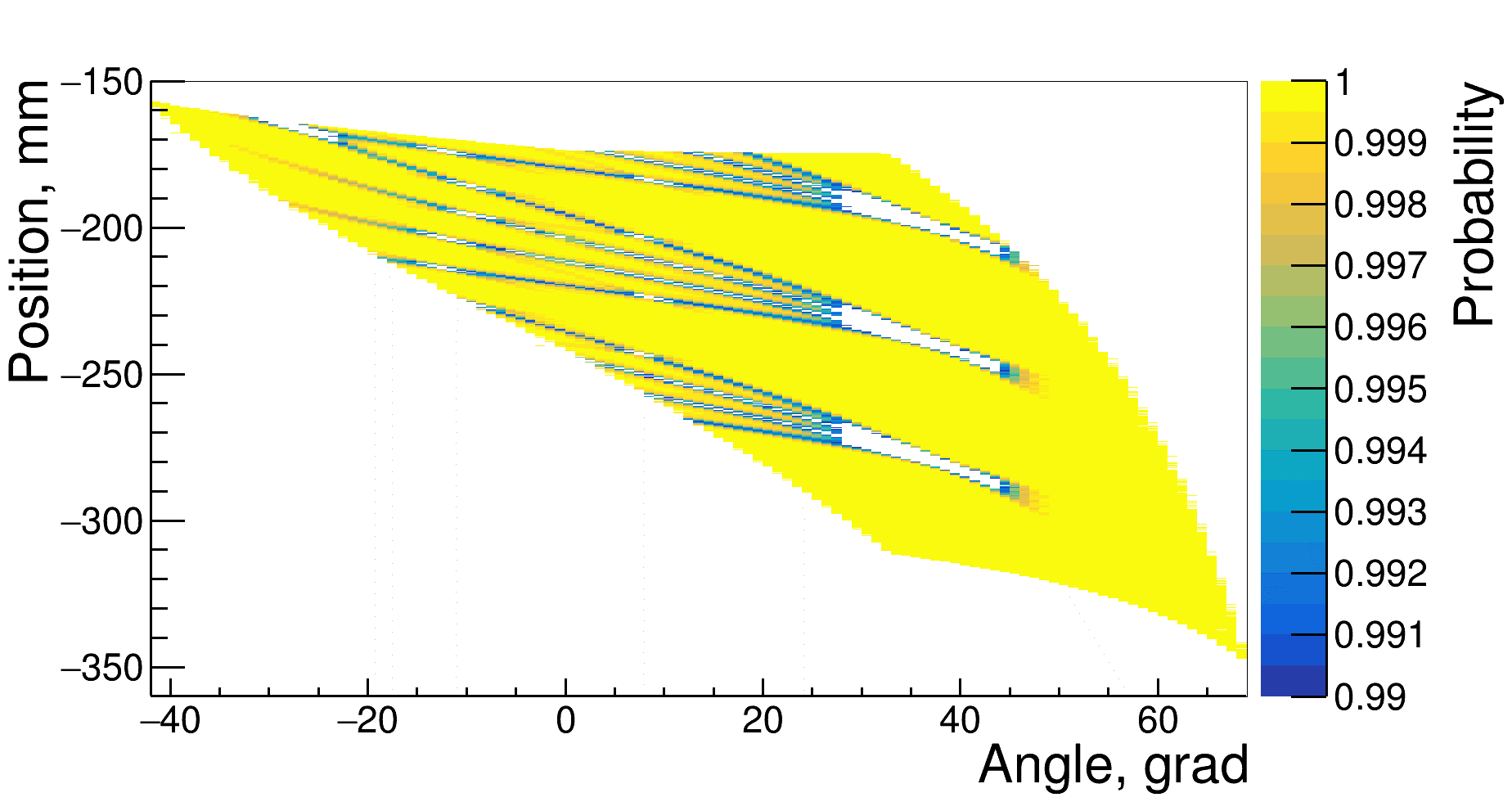}}
	\caption{\small Registration probability map obtained by simulation:
		with scale from 60\% to 100\% (a). Same plot with “magnifier”: scale is from 99\% to 100\% (b)}
	\label{fig: crv4x4effmap}
\end{figure}

At first, the registration probability map was created (Fig. \ref{fig: crv4x4effmap}). Then, using real angular distribution of the cosmic muons, the spot overall efficiency of the module was calculated.

Was found that by simulation, that the overall module efficiency for 4x4 CRV module is 99.74\% for average light yield of 21 ph.e and threshold set on 5 ph.e. level. This result is close to the result obtained by direct experiment.

We also calculated with Geant the overall efficiency for the light yield of 25 ph.e. but with same other conditions. Was found that the CRV module efficiency increased to 99.93\%. This result shows that it is necessary to improve the light yield from strip to get highly efficient system required for CRV.

%% file: parts/Conclusions.tex
A Simplified Light Yield Distribution (SLYD) method utilizing the transverse distribution of light yield to simulate the CRV module  efficiency is developed. Transversal scan using $^{90}Sr/^{90}Y$
$\beta-$source to mimic cosmic muon introduced. The strips with 40x7 $mm^2$ and 40x10 $mm^2$ in cross-section were scanned to provide the required for this method data. Both strips were equipped with 1 WLS fiber glued into the central groove.

Using GEANT4, the registration probability maps using SLYD method were created for 15x4 (4 layer by 15 strips on each layer) CRV modules for different patterns (layers shifts to each other array) to find best pattern. Simulation shows that the overall registration efficiency depends on patterns, strips light yield and geometrical parameters like gap between strips for neighbor one. The simulation was done for strips with cross-section of 40x7 and 40x10 $mm^2$.

Was found that efficiency for the CRV module based on 40x7 $mm^2$ in cross-section strips with average light yield set to 21 ph.e. based on real data for such strips and threshold level on 5 ph.e is less than 99.93\% for the best pattern. And it will drop to 99.67\% in 3-year term and to 83.17\% for 13-year term in base of the aging rate at 6\% per year. 

Efficiency for similar CRV module but based on 40x10 $mm^2$ in cross-section strips with average 30 light yield was simulated. Was found that with same conditions: almost 100\% efficiency in initial, 99.999\% in 3-year term and 99.98 in 13-years term. Better "strip thickness VS gap" ratio with the other same conditions – 20 (10/0.5) for 10-mm-thick strip against 14 (7/0.5) for 7-mm-thick strip and, in addition, 10-mm-thick strip has by 30\% more light - these two factors help to achieve these results.

The 4x4 CRV module using 7x40x3000 $mm^2$ strips was created. The efficiency at 2500 mm distance from SiPMs was found using cosmic muons. Th average light yield at this position was 21 ph.e. Also, the simulation for efficiency using SLYD method in GEANT4 of such module was provided for comparison with experimental one. Was found that overall efficiency for this module at this position was at 99.69\% level and the simulation results for this module were 99.74\% which is in good agreement with experimental data.

It is important to note that for the systems required the high level registration efficiency at the 99.99\% and more, it is important to improve the light yield as much as possible and achieve the gap between neighbor scintillation volumes as small as possible.

%% file: parts/Acknowledgments.tex
The authors would like to express deepest appreciation to V.Kolomoec, V.Rogozin and I.Prokhorov who provided a high level of technical support on each step of this research.

%% file: HighEffMuonRegSys.bbl
\begin{thebibliography}{10}
\expandafter\ifx\csname url\endcsname\relax
  \def\url#1{\texttt{#1}}\fi
\expandafter\ifx\csname urlprefix\endcsname\relax\def\urlprefix{URL }\fi
\expandafter\ifx\csname href\endcsname\relax
  \def\href#1#2{#2} \def\path#1{#1}\fi

\bibitem{mu2e_tdr2014}
L.~Bartoszek, et~al., {Mu2e Technical Design Report}, Tech. rep., FERMILAB (10 2014).
\newblock \href {http://arxiv.org/abs/1501.05241} {\path{arXiv:1501.05241}}, \href {http://dx.doi.org/10.2172/1172555} {\path{doi:10.2172/1172555}}.

\bibitem{comet_tdr2018}
R.~Abramishvili, et~al., {COMET Phase-I Technical Design Report}, Tech. Rep.~3, KEK/J-PARC (2020).
\newblock \href {http://arxiv.org/abs/1812.09018} {\path{arXiv:1812.09018}}, \href {http://dx.doi.org/10.1093/ptep/ptz125} {\path{doi:10.1093/ptep/ptz125}}.

\bibitem{poissdiss}
G.~L. Squires, Practical Physics, Cambridge University Press, 2001.
\newblock \href {http://dx.doi.org/10.1017/cbo9781139164498} {\path{doi:10.1017/cbo9781139164498}}.

\bibitem{Giant4}
{Geant4 Collaboration}, {Geant4. A simalution toolkit}, available at \url{http://geant4.web.cern.ch/}, (accessed on 09/23/2023).

\bibitem{kuraraySCSF81}
{Kuraray Corp., Japan}, Scintillating fibers, available at \url{http://kuraraypsf.jp/psf/sf.html}, (accessed on 23/09/2023).

\bibitem{uniplast}
{OOO Uniplast, Vladimir, Russia}, Development and production of plastic scintillators, \url{http://uniplast-vladimir.com/}, (accessed on 23/09/2023, in Russian).

\bibitem{kurarayY11}
{Kuraray Corp., Japan}, Wavelength shifting fibers, available at \url{http://kuraraypsf.jp/psf/ws.html}, (accessed on 23/09/2023).

\bibitem{caenDT5702}
{CAEN SpA, Italy}, Model dt5702, 32 channel sipm readout board for cosmic rays veto boxed, available at \url{https://www.caen.it/products/dt5702/}, (accessed on 23/09/2023).

\bibitem{hamamatsuS13360}
{Hamamatsu Photonics K.K., Japan}, {MPPC (Multi-Pixel Photon Counter) S13360 series}, available at \url{https://www.hamamatsu.com/content/dam/hamamatsu-photonics/sites/documents/99_SALES_LIBRARY/ssd/s13360_series_kapd1052e.pdf}, (accessed on 23/09/2023).

\bibitem{emi9814}
{ET Enterprises}, {9814B series data sheet}, available at \url{https://my.et-enterprises.com/images/data_sheets/9814B.pdf}, (accessed on 23/09/2023).

\bibitem{picoamK6487}
{Keithley Products, TEKTRONIX INC, United States}, {Keithley 6487 picoammeter}, available at \url{https://www.tek.com/en/datasheet/series-6400-picoammeters/6487-picoammeter-voltage-source}, (accessed on 23/09/2023).

\bibitem{scintCDFaging}
A.~Artikov, D.~Chokheli, G.~Pauletta, A.~Simonenko, {The loss of light yield with time in the CDF II scintillation counters}, Nucl. Instrum. Methods Phys. Res. A 672 (2012) 46 -- 51.
\newblock \href {http://dx.doi.org/10.1016/j.nima.2011.12.112} {\path{doi:10.1016/j.nima.2011.12.112}}.

\bibitem{hvsys}
{HVSys, JINR, Russia}, {Calibrated LED sources of light flashes}, available at \url{http://hvsys.ru/images/data/news/10_small_1368803142.pdf }, (accessed on 23/09/2023).

\end{thebibliography}
